\documentclass[twocolumn]{aastex631}

\usepackage{hyperref}
\usepackage{graphicx}
\usepackage{subfigure}
\usepackage{dcolumn}
\usepackage{bm}
\usepackage{float}
\usepackage{amsmath}
\usepackage{natbib}
\usepackage{courier}
\usepackage{soul}
\usepackage{longtable}

\begin{document}
\title{Asteroseismology of Metal-Poor Red Giants Observed by TESS}

\author{Corin Marasco}
\affiliation{Department of Astronomy, University of Florida, Gainesville, FL 32611, USA}

\author{Jamie Tayar}
\affiliation{Department of Astronomy, University of Florida, Gainesville, FL 32611, USA}

\author{David Nidever}
\affiliation{Department of Physics, Montana State University, Bozeman, MT 59717, USA}

\begin{abstract}
Galactic archeology has long been limited by a lack of precise masses and ages for metal-poor stars in the Milky Way's thick disk. However, with TESS providing a growing number of photometric observations, it is possible to calculate masses and ages for more solar-like oscillators than ever using asteroseismology. We have used the \texttt{pySYD} pipeline to determine global asteroseismic parameters and calculated the masses and ages of $506$ metal-poor ([M/H] $<$ -0.5) red giants observed by TESS. Our findings appear to show metallicity-dependent mass loss on the upper red giant branch and identify a set of ``young" high-$\alpha$ stars that have been detected in other studies. We also find that $32.6\%$ of the metal-poor stars appear to be binary interaction products and four stars with stellar ages that could be from the Gaia Enceladus/Sausage system. In combination with existing ages from Kepler/K2, this data can be compared to galactic evolution models to better determine the formation history of the galaxy.

\end{abstract}

\section{Introduction} \label{sec:intro}

Galactic archeology examines the characteristics of stars in our galaxy to understand the formation of the Milky Way. Remnants of the galaxy's history, in the form of old stars, contain crucial information about the past, which can be used to constrain galactic evolution models \citep{miglio2009,sharma2016,silva2018,warfield2024}. Through these studies, galactic archeology connects information about the Milky Way's past and present to construct a comprehensive narrative of how it came to be.

Galactic archeology relies on accurate measurements of stellar ages, abundances, and kinematics for a large population of stars throughout the galaxy. Metal-poor stars are of particular interest due to their older ages representing a time when the galactic disk was still forming, before Type Ia supernovae had enriched the interstellar medium with iron and heavier elements \citep{timmes1995}. This critical stage in galactic evolution is currently not well-understood. The age of the Milky Way's thick disk is debatable, with estimates ranging from $7$ Gyrs to $11$ Gyrs, and it is unclear how the bimodal $\alpha$-abundance in the thin and thick disk developed \citep{lian2020, lu2024}. Metal-poor stars were formed at critical points in the galaxy's history, evidenced by the fact that the $\alpha$ bimodality is most distinct in the low-metallicity regime. Therefore, if we could precisely determine their ages, these stars could fill in gaps in our understanding of the Milky Way's evolution.


Asteroseismology, or the study of stellar oscillations, is being used to determine the masses and ages of large numbers of distant and single stars, and has therefore been used to advance galactic archeology in recent years \citep{bovy2014, epstein2014, silva2018, marc2018, hon2021, apok2}. Red giants are typically used for those measurements, as their bright luminosities make them easier to see, leading to more accurate spectroscopy and photometry. Additionally, they are solar-like oscillators, which allows us to use the more well-understood asteroseismic scaling relations to calculate their masses and radii \citep{ulrich_1986,brown_1991,asteroseismology_eqns}.

In the past two decades, asteroseismology has been rapidly advancing due to long duration photometric observations of stars from space telescopes. The CoRoT \citep{corot}, Kepler \citep{kepler}, K2 \citep{k2}, and most recently TESS \citep{tess} missions have provided long duration, precise observations for an unprecedented amount of stars in the Milky Way. These observations have helped refine stellar models and therefore improve asteroseismic calculations of stellar masses and ages \citep{stello2016,tayar2017,joyce2018,warfield2024}.

Before the recent launch of TESS, asteroseismic observations were limited to the small fields of view provided by the CoRoT, Kepler, and K2 missions. TESS observes nearly the entire sky, opening up hundreds of thousands of oscillating red giants to asteroseismic analysis \citep{hon2021}. TESS has not yet observed for long enough to perform asteroseismic analysis on all the stars in its field of view. However, after all parts of the sky have been observed for multiple sectors over the coming years, we expect that it could eventually provide precise asteroseismic ages for stars distributed across the Milky Way. With those ages, galactic archeology would have a revolutionary amount data with which to piece together the galaxy's history.

TESS gathers observations by dividing the sky into $26$ sections and observing each section for $27$ days at a time \citep{tess}. The downside to observing the entire sky in sections is that TESS collects less data per star than similar missions such as Kepler, leading to lower frequency resolution for the oscillations. However, over the duration of TESS's lifetime, it has now observed each section for at least $54$ days, with the Southern continuous viewing zone, where the sectors overlap the most, gathering a total of over $900$ days of observations. With that amount of data, it is possible to try and perform asteroseismology on red giants in a wider field of view than ever before using TESS light curves.

Recent studies have successfully performed asteroseismology on TESS data, with masses determined to $5$-$10\%$ precision and ages to $20\%$ \citep{silva2020,mackereth2021,hon2021,huber2022,stello2022,hatt2023,zhang2024,lindsay2024}. These studies have typically focused on either large surveys or stars with solar-like metallicities, often neglecting metallicity-dependent corrections. Therefore, asteroseismic parameters have not been explicitly determined for more than a handful of red giants at low metallicities in the TESS field of view \citep{chaplin2020,huber2024}. A major reason for this is TESS's short observational periods. Giant stars are visible from greater distances, yielding a larger stellar sample, but short observations from TESS lead to worse resolution on the low-frequency end of the power spectrum, where oscillations for metal-poor giants typically occur. However, as TESS continues to gather more observations of metal-poor stars, their oscillations are becoming more defined, leading to more accurate asteroseismic analysis. The other complication with calculating the masses of low-metallicity stars is that, with Kepler and K2 data, they have notoriously yielded higher asteroseismic masses than expected for old stars, even after implementing corrections for the low-metallicity regime \citep{epstein2014,apok2}
. Therefore, although the metal-poor red giants in the much wider TESS field of view are a historically significant set of stars, the combination of noisy TESS data and previous issues with the asteroseismology of metal-poor stars have dissuaded detailed asteroseismology of this set of
 stars.

This study addresses that gap in knowledge by using TESS data to asteroseismically determine the masses and ages of old, metal-poor stars over the broadest possible field of view to date. We then relate those asteroseismic results to the abundances and kinematics of the stellar population to compare the ages of stars in the thin and thick disk and identify stars that are possibly from an accreted dwarf galaxy.


\section{The Data}


\subsection{Sample Selection}
The asteroseismic data for this study was provided by TESS \citep{tess}. TESS divides the sky into sectors, each of which is observed for $27$ days at a time, with a $4$-hour gap at $13.7$ days to downlink data. Because TESS has now completed five years of observations, most regions of the sky have been observed for at least two $27$-day periods. However, because the segments overlap with each other at the ecliptic poles, some stars have many more than two sectors of data. In this study, we found that approximately $5$ or more sectors of data were needed to obtain a frequency resolution where we could correctly identify a metal-poor giant's asteroseismic parameters. This limited us to giants in or near the TESS continuous viewing zones, where sectors overlapped and more data was available.

The list of possible stars to study was determined first by cross-referencing a list of oscillating red giants from TESS---compiled using machine learning in \citet{hon2021}---with the Apache Point Observatory Galactic Evolution Experiment (APOGEE) survey to determine their metallicities \citep{apogee2017}. This left us with a list of 15,000 red giants that TESS has detected oscillations in and have spectroscopically-determined characteristics available in APOGEE \citep{artemis2023}.

The APOGEE survey is a spectroscopic study of over 650,000 red giants in the Milky Way galaxy with the goal of understanding the formation of the Milky Way with detailed studies of ancient stars. We used observational data analyzed in Data Release 17 of APOGEE  \citep{dr17}, which was obtained during the fourth stage of the Sloan Digital Sky Survey \citep{sdss4}. APOGEE uses a high-resolution (R $\sim$ $22,500$) infrared spectrograph \citep{spectrographs} and is mounted on both the Sloan Foundation 2.5m telescope \citep{north_tele} at Apache Point Observatory, New Mexico and the 2.5m Ir\'en\'ee DuPont telescope \citep{south_tele} at Las Campanas Observatory, Chile.

The spectroscopic parameters in APOGEE were calculated using the APOGEE Stellar Parameters and Chemical Abundances Pipeline (ASPCAP). ASPCAP determines stellar parameters, such as T$_\mathrm{eff}$, log$g$, metallicity, and abundances, by fitting the APOGEE spectrum and comparing it to synthetic spectra \citep{aspcap2016}. These synthetic spectra are matched to observations based on atmospheric model grids derived in \citet{holtzman2018} and corrected in \citet{jonsson2020}.

Gaia Data Release 3, which includes astrometry and photometry information for over 1 billion sources, was used to collect kinematic parameters and radius estimates of the stars \citep{gaia3}.

\begin{figure}[H]
\centering\includegraphics[scale=.48]{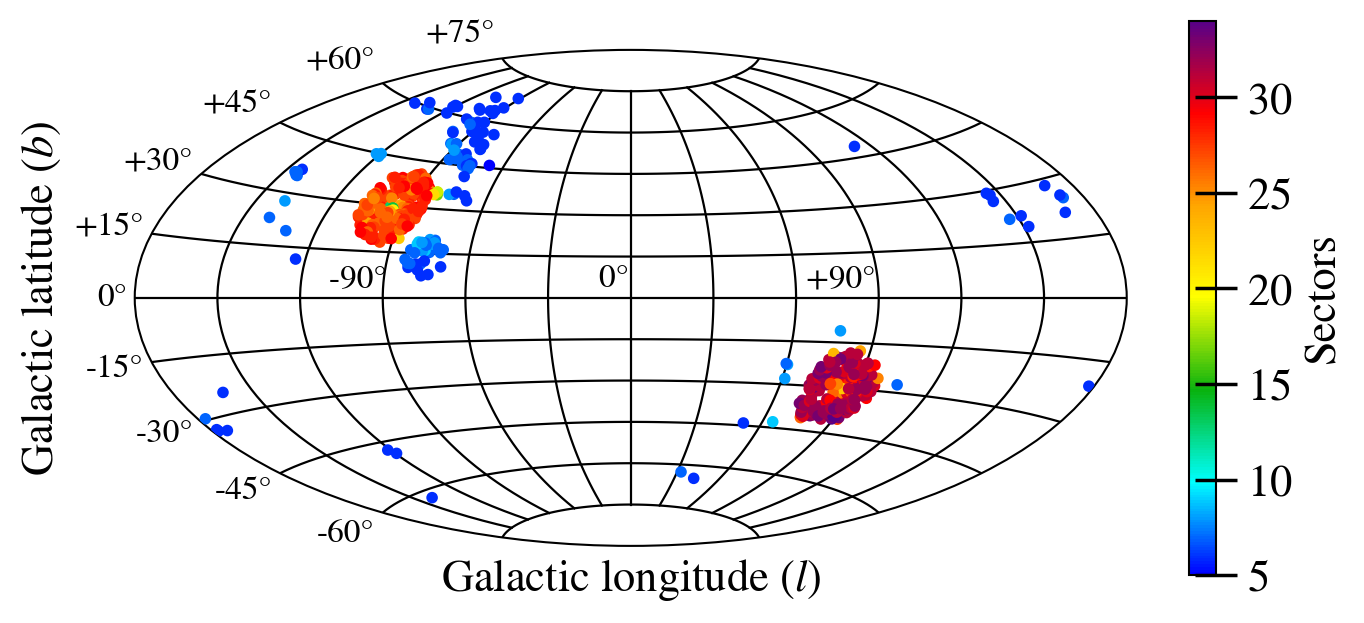}
\caption{\label{fig:dis_l} The galactic distribution of all the low-metallicity stars in \citet{hon2021} and APOGEE DR17 with more than $5$ sectors of data available from TESS. The number of TESS sectors that each star was observed for is indicated by its color. The orange clump is TESS's northern continuous viewing zone (NCVZ), and the red clump is the southern continuous viewing zone (SCVZ). The SCVZ stars have more sectors available because TESS has observed the southern hemisphere three times and is on its third observation of the northern hemisphere (as of June 2024).}
\end{figure}

From that cross-referenced list, we selected all stars with a metallicity of [M/H] $< -0.5$ according to APOGEE data. This list was then filtered so that only stars with more than $5$ sectors of TESS data (from TESS Cycles 1-5) were included. Unfortunately, as shown in Figure \ref{fig:dis_l}, this last filter limited us to stars near or in TESS's two CVZs, where sectors overlapped and targets were observed more than once in a cycle. This left $515$ metal-poor red giants with enough data to calculate their masses using asteroseismic methods.

\subsection{Light Curves}
The light curves for each star were downloaded from MIT's Quick-Look Pipeline (QLP, \citealt{qlp1,qlp2})
using the Python package \texttt{Lightkurve} \citep{lightkurve}. The first and last days of each sector's light curve were removed to ensure that instrumental irregularities during those times did not affect the asteroseismology \citep{avallone2022}. Each sector was normalized and smoothed on a timescale of a few days, and outliers of $>2.75\sigma$ were removed. After stitching all of the observations together, the light curve was normalized again.

Gaps in TESS light curves can last for years between TESS missions. When left untreated or filled with interpolated data, as suggested in \citet{bedding2022} and \citet{garcia2014}, respectively, we found that these led to the \texttt{pySYD} pipeline either failing to converge or calculating asteroseismic parameters that were further from the expected values. Therefore, we used an approach similar to \citet{gaps} and shifted the time stamps to remove gaps greater than $10$ days.

Additionally, we removed Sector $8$ from every light curve, due to irregular variations detected in that sector that affected the asteroseismology of most stars. During data collection for sector $8$, TESS was temporarily deactivated for $3$ days by an instrument anomaly and the camera temperature was increased for a short period, which likely caused rapid variations in flux, interfering with the asteroseismology \citep{tess_notes}.

We also experimented with removing additional sectors that had fluxes with unusually large amplitudes or strange oscillations. However that appeared to have a slight negative effect on the accuracy of the asteroseismic analysis, so, other than sector $8$, the entire light curve was used.

\subsection{Asteroseismology}

In order to calculate the masses of the stars, the oscillation properties $\Delta\nu$ and $\nu_\textrm{max}$ were derived for every star through the features of its power density spectrum. The large frequency separation is represented by $\Delta\nu$ and describes the difference in frequency between the radial oscillation modes. This quantity scales with the square root of the mean density of a star and is approximately equal to the inverse of the sound travel time through a star \citep{brown_1991,ulrich_1986}. The other value, $\nu_\textrm{max}$, is the frequency at which a star's power spectrum reaches its maximum and is related to the atmospheric pressure of a star, and therefore its surface gravity \citep{asteroseismology_eqns}.

These asteroseismic values can be used to calculate the mass, radius, and log$g$ of a star using the following equations from \citet{asteroseismology_eqns}.

\begin{equation}\label{eqn:mass}
    \frac{M}{M_\odot}=
    \left(\frac{\nu_\textrm{max}}{\nu_{\textrm{max,}\odot}}\right)^3
    \left(\frac{\Delta\nu}{\Delta\nu_\odot}\right)^{-4}
    \left(\frac{T_\textrm{eff}}{T_{\textrm{eff,}\odot}}\right)^{3/2}
\end{equation}

\begin{equation}\label{eqn:radius}
    \frac{R}{R_\odot}=
    \left(\frac{\nu_\textrm{max}}{\nu_{\textrm{max,}\odot}}\right)
    \left(\frac{\Delta\nu}{\Delta\nu_\odot}\right)^{-2}
    \left(\frac{T_\textrm{eff}}{T_{\textrm{eff,}\odot}}\right)^{1/2}
\end{equation}

\begin{equation}\label{eqn:logg}
    \frac{g}{g_\odot} =\frac{\nu_\textrm{max}}{\nu_{\textrm{max,}\odot}}\left(\frac{T_\textrm{eff}}{T_{\textrm{eff,}\odot}}\right)^{1/2}
\end{equation}

\begin{figure}[H]
\begin{subfigure}
    \centering\includegraphics[scale=.55]{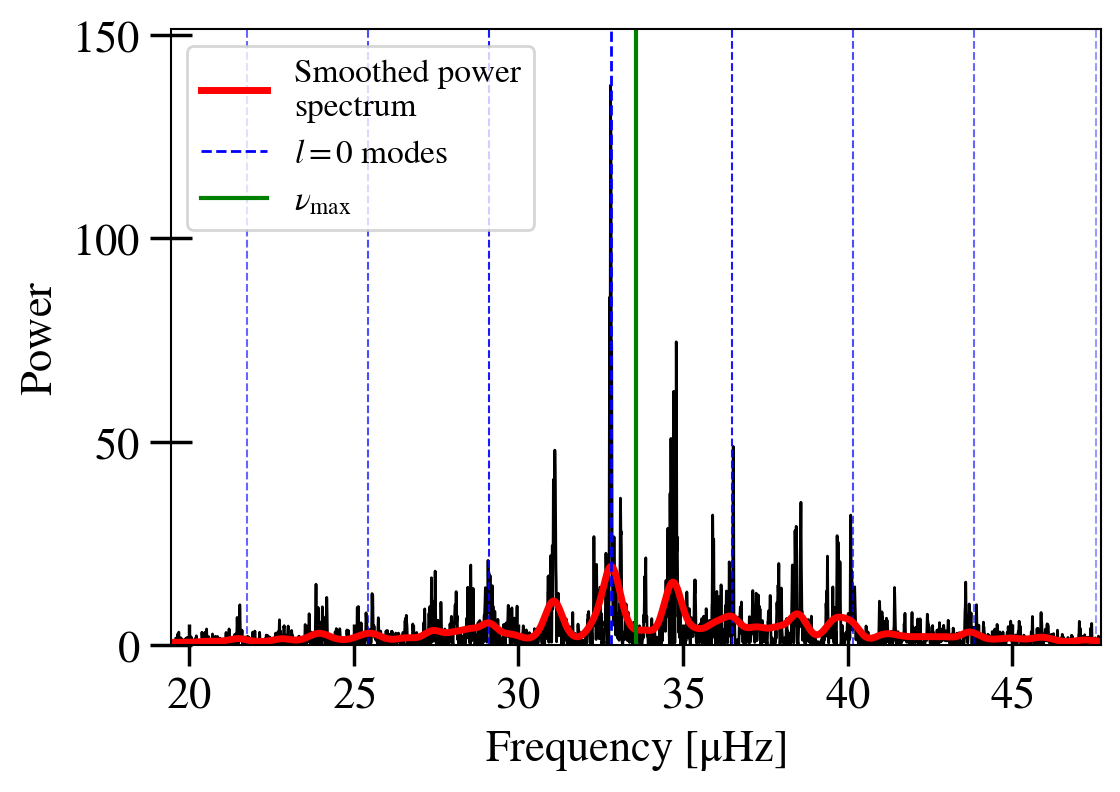}
\end{subfigure}

\begin{subfigure}
    \centering\includegraphics[scale=.5]{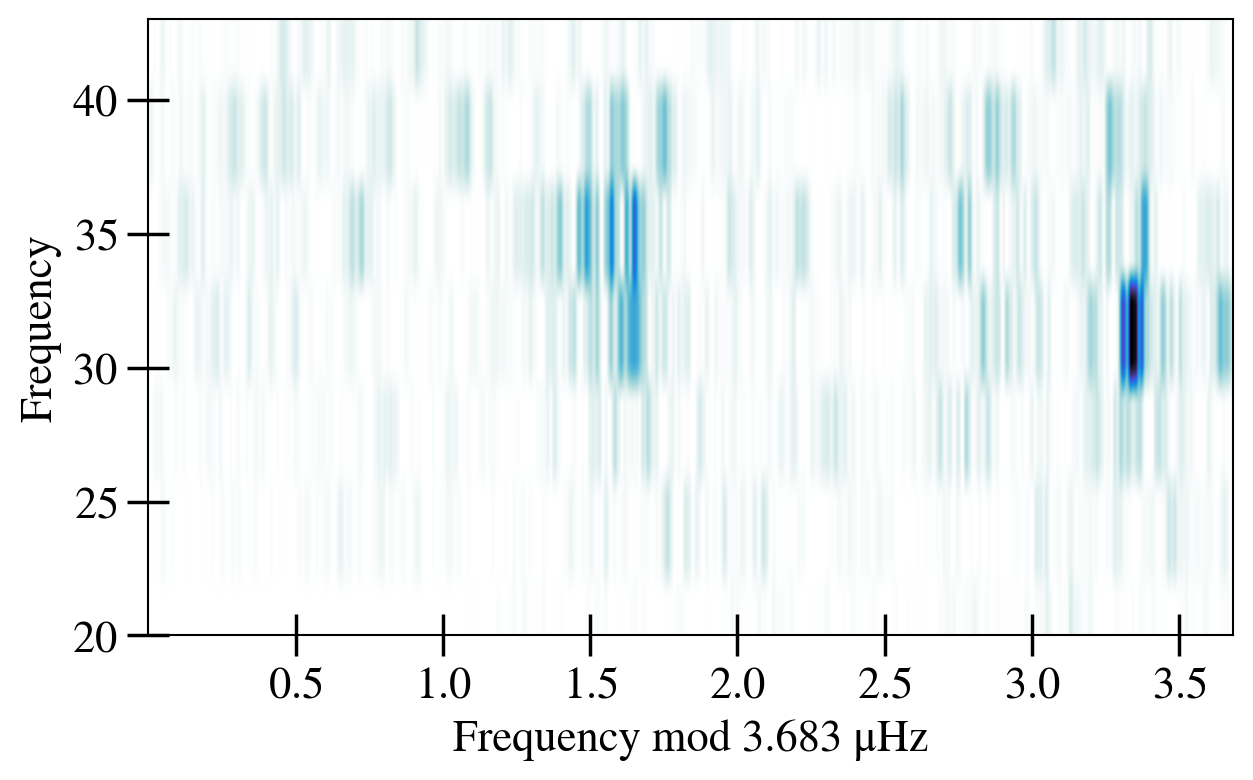}
\end{subfigure}
\caption{\label{fig:astero} The top figure shows the power spectrum of the red giant TIC 350561656. The star's $\nu_\textrm{max}$, as identified by \texttt{pySYD}, is shown with a solid green line. The $l=0$ modes are shown with the dashed blue lines, which are offset by $\Delta\nu=3.683$ to show the peaks in the spectrum. The red curve shows the power spectrum after heavy smoothing was applied. The echelle diagram in the bottom figure shows the power spectrum split up by $\Delta\nu=3.683$ and stacked vertically, showing that the peaks line up with each other and revealing an $l=1$ mode on the left and a faint $l=2$ and strong $l=0$ mode on the right.}
\end{figure}

To perform the asteroseismology, the \texttt{pySYD} pipeline--a Python version of the \texttt{SYD} pipeline--was used \citep{syd,pysyd}. The pipeline calculated the background subtracted power spectrum, $\nu_\textrm{max}$, and $\Delta\nu$ for each light curve. Each power spectrum was visually inspected to confirm the $\nu_\textrm{max}$ and $\Delta\nu$ results from \texttt{pySYD}, and many of the $\Delta\nu$ values required some additional adjustments. Therefore, the Echelle diagram and power density spectrum of each star created by \texttt{pySYD} was examined to ensure that $\Delta\nu$ matched the average large frequency spacing, and was adjusted if necessary \citep{echelle}. An example of a typical power spectrum is shown in Figure \ref{fig:astero}.


\subsection{Evolutionary States}
Before calculating the masses of the red giants in our set, some corrections needed to be applied to the $\Delta\nu$ and $\nu_\textrm{max}$ values. These corrections vary depending on the star's evolutionary state, which needed to be estimated for the stars in our set. The APOGEE-Kepler Asteroseismic Science Consortium (APOKASC-3) catalog \citep{apokasc3} contains the asteroseismic evolutionary states of many similar stars, so we used that data as a training set to predict the evolutionary states of the low-metallicity TESS red giants using machine learning.

We assumed that the giants in our set of low-metallicity stars were either in the red giant branch (RGB) or red clump (RC) phase of their evolution. Therefore, only the RGB and RC stars in APOKASC were used in our training set to estimate evolutionary states. Another cut was then made so that only the lower-metallicity APOKASC stars ([M/H] $<-0.3$) were used in the training set. We found that the $-0.5$ metallicity cutoff used with the APOGEE stars was too strict for the training set and did not provide a sufficient sample.

Core He-burning RC stars are hotter than shell-burning RGB stars with the same mass and composition. That temperature offset between the He-burning phase and the first ascent red giant branch is metallicity-dependent, and this separation is larger for low-metallicity stars \citep{giriardi1999}. For this reason, we estimate the temperature cutoff between evolutionary states for each star ($T^{\textrm{co}}_{\textrm{eff}}$) using the equation proposed by \citet{bovy2014}.

\begin{equation}
    T^{\textrm{co}}_{\textrm{eff}} = \frac{\textrm{log\ }g - 2.5}{0.0018\textrm{\ } \tfrac{\textrm{dex}}{\textrm{K}}} + -382.5\textrm{\ }\tfrac{\textrm{K}}{\textrm{dex}} \textrm{\ [Fe/H]}+4607\textrm{\ K}
\end{equation}

\begin{figure}[H]
\centering\includegraphics[scale=.4]{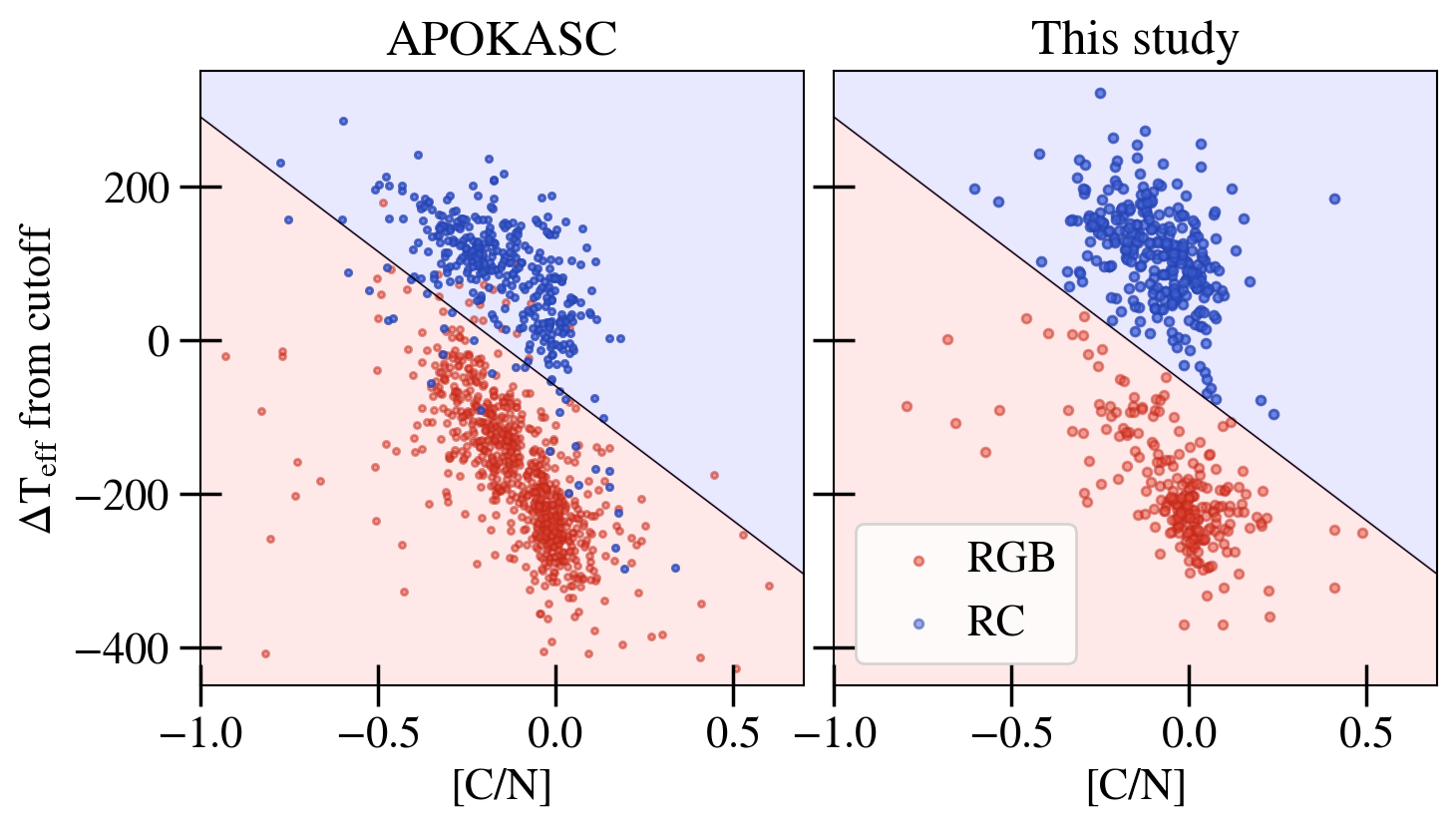}
\caption{\label{fig:ev_states} The left figure shows the SVM model fit to the APOKASC data to estimate a distinguishing line between RGB and RC stars. The plot on the right shows the stars from our study with the SVM fit used to determine evolutionary states.}
\end{figure}

The temperature difference between $T^{\textrm{co}}_{\textrm{eff}}$ and a star's effective temperature was calculated to separate the RC and RGB stars. The $\textrm{[C/N]}$ ratio was also considered, as it can be used as a proxy for the mass, temperature, metallicity, and log$g$ of each red giant \citep{martig2016}. When the difference between a star's temperature and cutoff temperature ($\Delta T_\textrm{cutoff}=  T_{\textrm{eff}}-T^{\textrm{co}}_{\textrm{eff}}$) were plotted with the $\textrm{[C/N]}$ ratio for the APOKASC-3 stars, it showed a clear separation between evolutionary states (Figure \ref{fig:ev_states}).

This separation was fitted with a linear support vector machine (SVM) model using \texttt{scikit-learn} \citep{sklearn}. The low-metallicity APOKASC-3 stars were used as training data to determine approximately where the separation between evolutionary states was. This model estimated with $94.8\%$ accuracy that stars with $\Delta T_\textrm{cutoff}>-299\textrm{[C/N]}-103.5$ were RC stars and stars below that line were in the RGB stage. That model was then applied to our low-metallicity TESS stars, and produced what appeared to be a reliable estimate of the evolutionary states (see Figure \ref{fig:ev_states}).

\subsection{Pipeline Corrections}
With the evolutionary states determined, we calculated the corrections to $\Delta \nu$ using the \texttt{asfgrid} code written by \citet{asfgrid} and updated in \citet{asfgrid6}, which was constructed using stellar models of stars with lower metallicities ($-3<\textrm{[Fe/H]}<0.4$). These corrections ($f_{\Delta \nu}$) were calculated for each star based on its evolutionary state, temperature, metallicity, and raw $\Delta\nu$ and $\nu_\textrm{max}$ values. The correction to $\nu_\textrm{max}$, $f_{\nu_\textrm{max}}$, was later calculated based on our seismic radii compared to the Gaia radii.


The $\Delta\nu$ and $\nu_\textrm{max}$ values were first adjusted based on a calibration study of other asteroseismology pipelines \citep{marc2018}. In lieu of calculating $\Delta\nu$ and $\nu_\textrm{max}$ using an average across multiple pipelines, the mean ratio of values from the \texttt{SYD} pipeline to the ensemble average, as presented in \citet{marc2018}, were used. \texttt{pySYD} is designed to be a Python version of the IDL-based \texttt{SYD} pipeline, so results from the two should closely match. The average offset between \texttt{SYD} and \texttt{pySYD} for $\nu_\textrm{max}$ was found to be $0.07\%$ and the $\Delta\nu$ offset was $0.004\%$ \citep{pysyd}. Therefore, the \texttt{SYD} corrections $X_\textrm{$\Delta\nu$}$ and $X_\textrm{$\nu_\textrm{max}$}$ were used to adjust the \texttt{pySYD} values to the assumed average among multiple pipelines.

The corrected $\Delta\nu$ and $\nu_\textrm{max}$ values were then calculated using the following equations:

\begin{equation}\label{eqn:numax_corrs}
    \nu_\textrm{max}=
    \left(\frac{\nu_\textrm{max, \texttt{pySYD}}}{X_{\nu_\textrm{max}}f_{\nu_\textrm{max}}}\right)
\end{equation}

\begin{equation}\label{eqn:dnu_corrs}
    \Delta\nu=
    \left(\frac{\Delta\nu_\textrm{\texttt{pySYD}}}{X_{\Delta\nu}f_{\Delta\nu}}\right)
\end{equation}

For RGB stars, $X_\textrm{$\Delta\nu$,RGB}=0.9995$ and $X_\textrm{$\nu_\textrm{max}$\textrm{,RGB}}=1.0006$ and for RC stars, $X_\textrm{$\Delta\nu$,RC}=1.0032$ and $X_\textrm{$\nu_\textrm{max}$\textrm{,RC}}=1.0010$. When calculating mass and radius, the average solar values across multiple pipelines, $\nu_{\textrm{max,}\odot}=3103.266$ and $\Delta\nu_\odot=135.146$, were used \citep{marc2018}.

\subsection{\texorpdfstring{$\nu_{\textrm{max}}$}{vmax} Calibration with Gaia Radii}

Before calculating masses, we derived a $f_{\nu_\textrm{max}}$ for Equation \ref{eqn:numax_corrs}. This was calculated based on our initial seismic radii calculations and the radii from \citet{hon2021} which were derived from Gaia DR2 data \citep{gaia2}. This allowed us to test the accuracy of the asteroseismology by comparing our $\Delta\nu$ and $\nu_\textrm{max}$ values against a radius that was determined independently.

\begin{figure}
\centering\includegraphics[scale=.45]{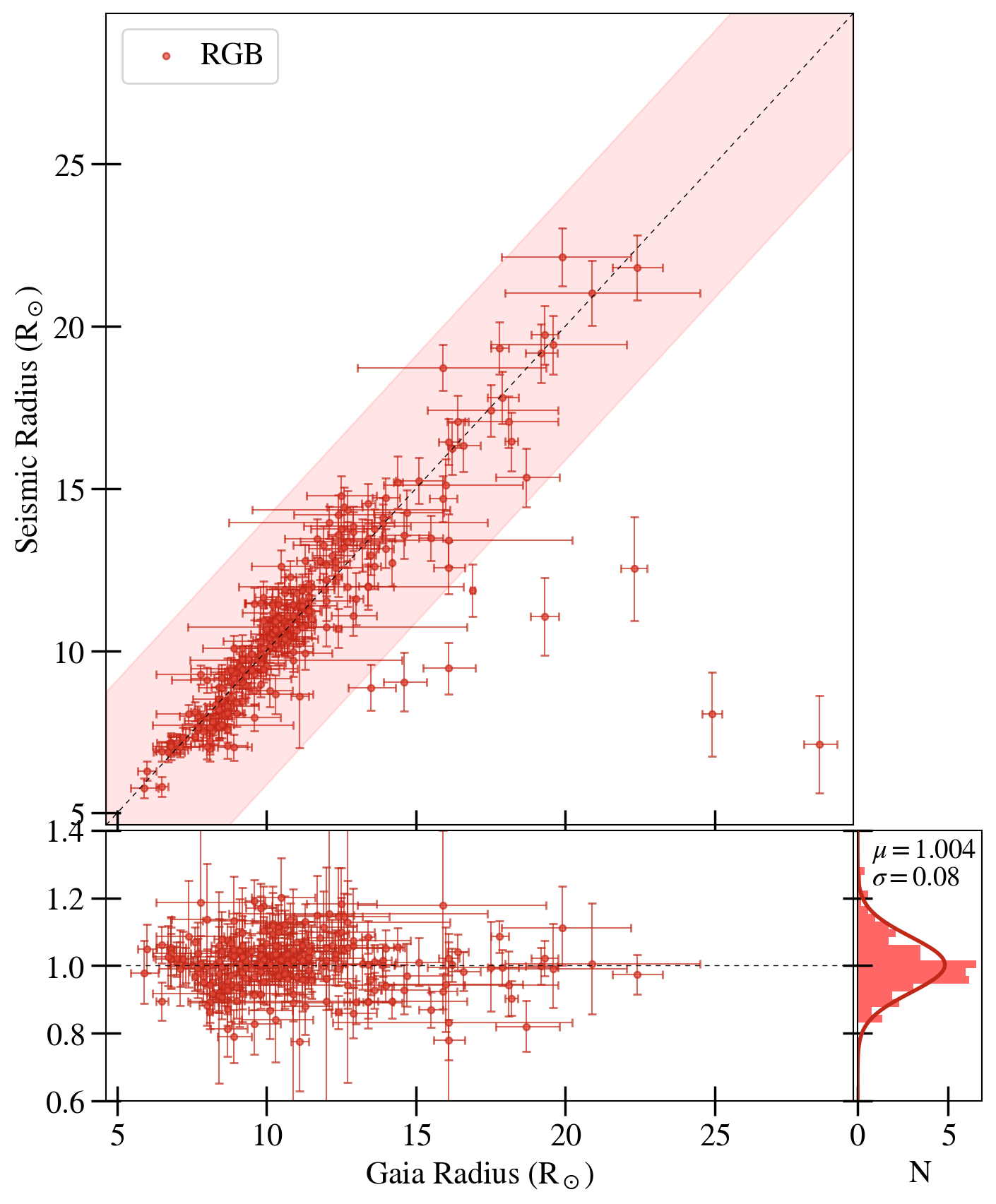}
\caption{\label{fig:rad_rgb} The top plot shows a one-to-one comparison of the radii of the RGB stars calculated from our asteroseismic parameters and the Gaia-derived radii from \citet{hon2021}. The broad red line stretches $2\sigma_\textrm{RGB}$ around the one-to-one line. The bottom plot shows the ratio between the two radii, after cutting out the outliers, which we defined as the stars outside of $2\sigma_\textrm{RGB}$. Without the outliers, the ratio has an average of $\mu_\textrm{RGB}=1.004$ and a scatter of $\sigma_\textrm{RGB}=0.080$.}
\end{figure}

\begin{figure}
\centering\includegraphics[scale=.45]{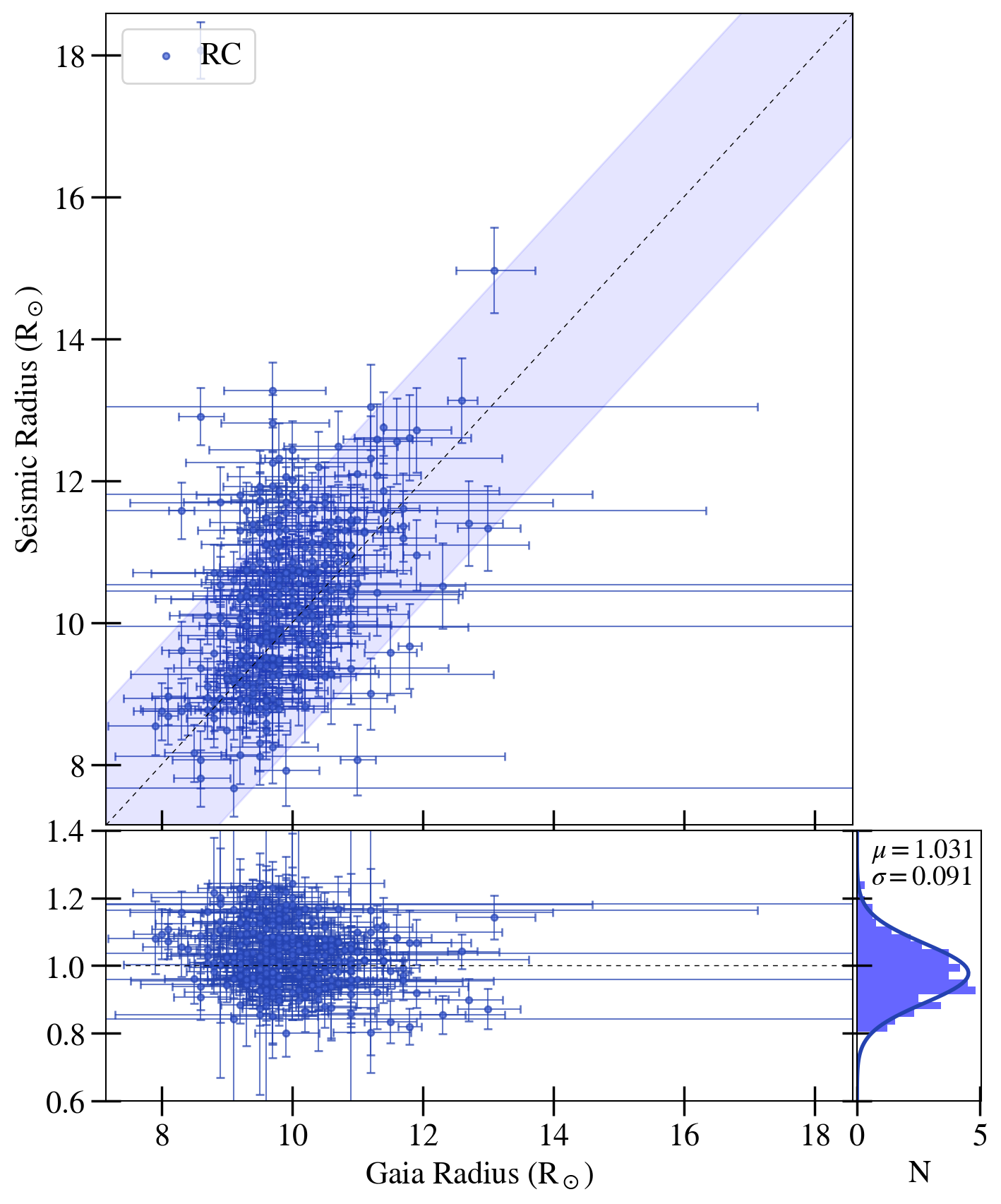}
\caption{\label{fig:rad_rc} The same as Figure \ref{fig:rad_rgb}, but with the RC stars instead of RGB. Here, the radius ratio has an average of $\mu_\textrm{RC}=1.031$ and a scatter of $\sigma_\textrm{RC}=0.091$.}
\end{figure}

The derived asteroseismic parameters, $\Delta\nu$ and $\nu_\textrm{max}$, for all of the stars were used in Equation \ref{eqn:radius} to calculate radii values (with $f_{\nu_\textrm{max}}=1$). The radii in \citet{hon2021} were calculated from luminosities determined using Gaia DR2 parallaxes, 2MASS temperatures and magnitudes, extinction values based on 3D dust maps \citep{bovy2016}, and bolometric corrections using MIST \citep{choi2016}.
The results are shown in Figure \ref{fig:rad_rgb} and \ref{fig:rad_rc}.

After removing the outliers, which we considered to be the RGB stars outside of $2\sigma_\textrm{RGB}=4.52\textrm{ R}_\odot$ and the RC stars outside of $2\sigma_\textrm{RC}=2.33\textrm{ R}_\odot$, we found a scatter of $0.080$ in the ratio between the asteroseismic and spectroscopic radii of RGB stars and a scatter of $0.091$ in the RC stars. We noted that these values were very similar to the average R$_\textrm{seis}$/R$_\textrm{Gaia}$ error in the ratio plots, which was $0.079$ for the RGB plot in Figure \ref{fig:rad_rgb} and $0.094$ for Figure \ref{fig:rad_rc}'s RC plot. This indicates that our error estimations on $\Delta\nu$ and $\nu_\textrm{max}$ (which were used to calculate R$_\textrm{seis}$ error) are consistent with errors derived empirically from the scalar difference between the two systems.

The average ratio between the two radii for the RGB and RC stars were $1.0037$ and $1.0311$ respectively. We used that offset to determine $f_{\nu_\textrm{max}}$, a correction ensuring that the average ratio of seismic to Gaia radii is $1$. The $f_{\nu_\textrm{max, RGB}}$ for every RGB star was set to $f_{\nu_\textrm{max, RGB}}=1.0037$ and every RC star was set to $f_{\nu_\textrm{max, RC}}=1.0311$. By adding the correction, we calibrate our $\Delta\nu$, $\nu_\textrm{max}$, and asteroseismic radii so they are consistent with observations, implying that the masses and ages calculated from those values will be reliable. While APOKASC-3 constructs $f_{\nu_\textrm{max}}$ as a function of radius, we find here that a simpler single value is appropriate for our data.

\begin{figure}
\centering\includegraphics[scale=.35]{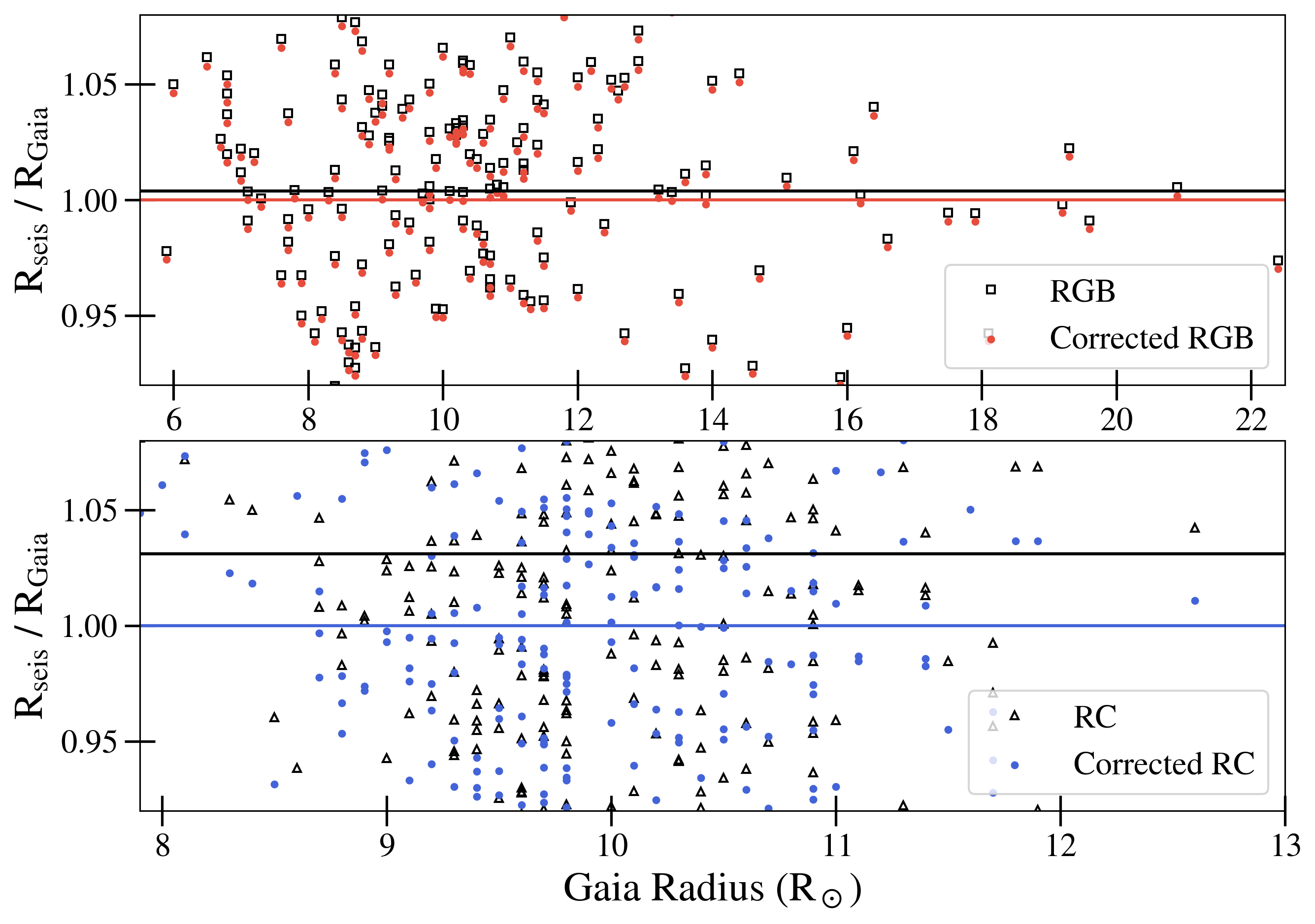}
\caption{\label{fig:rad_corr} Shows the ratio of seismic radii to Gaia-derived radii before and after applying $f_{\nu_\textrm{max}}$ to the seismic radius calculation. The horizontal lines show the average ratio of seismic to Gaia radii, demonstrating that after applying $f_{\nu_\textrm{max}}$, this average ratio is one. To demonstrate this shift, the plots have been cropped in the R$_\mathrm{seis}$/R$_\mathrm{Gaia}$ direction, however, all points that were not outliers in Figures \ref{fig:rad_rgb} and \ref{fig:rad_rc} were included in calculating the average.}
\end{figure}

The new seismic radii after applying $f_{\nu_\textrm{max}}$ are shown in Figure \ref{fig:rad_corr}. This $f_{\nu_\textrm{max}}$ term was also included in the mass calculations of Section \ref{sec:mass}.



\section{Analysis}

\subsection{Mass}\label{sec:mass}

\begin{figure*}
\begin{subfigure}
    \centering\includegraphics[scale=.392]{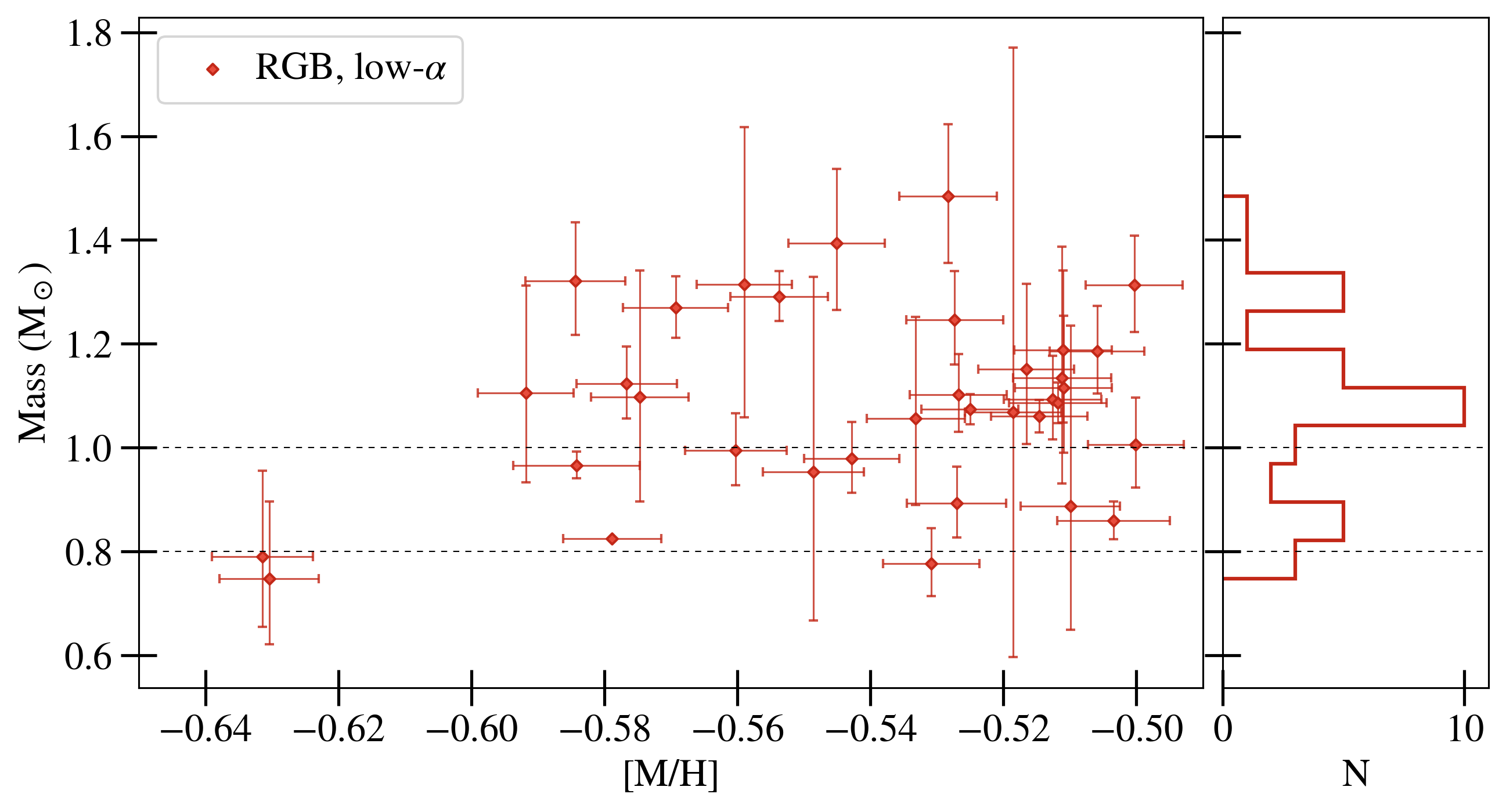}
\end{subfigure}
\begin{subfigure}
    \centering\includegraphics[scale=.392]{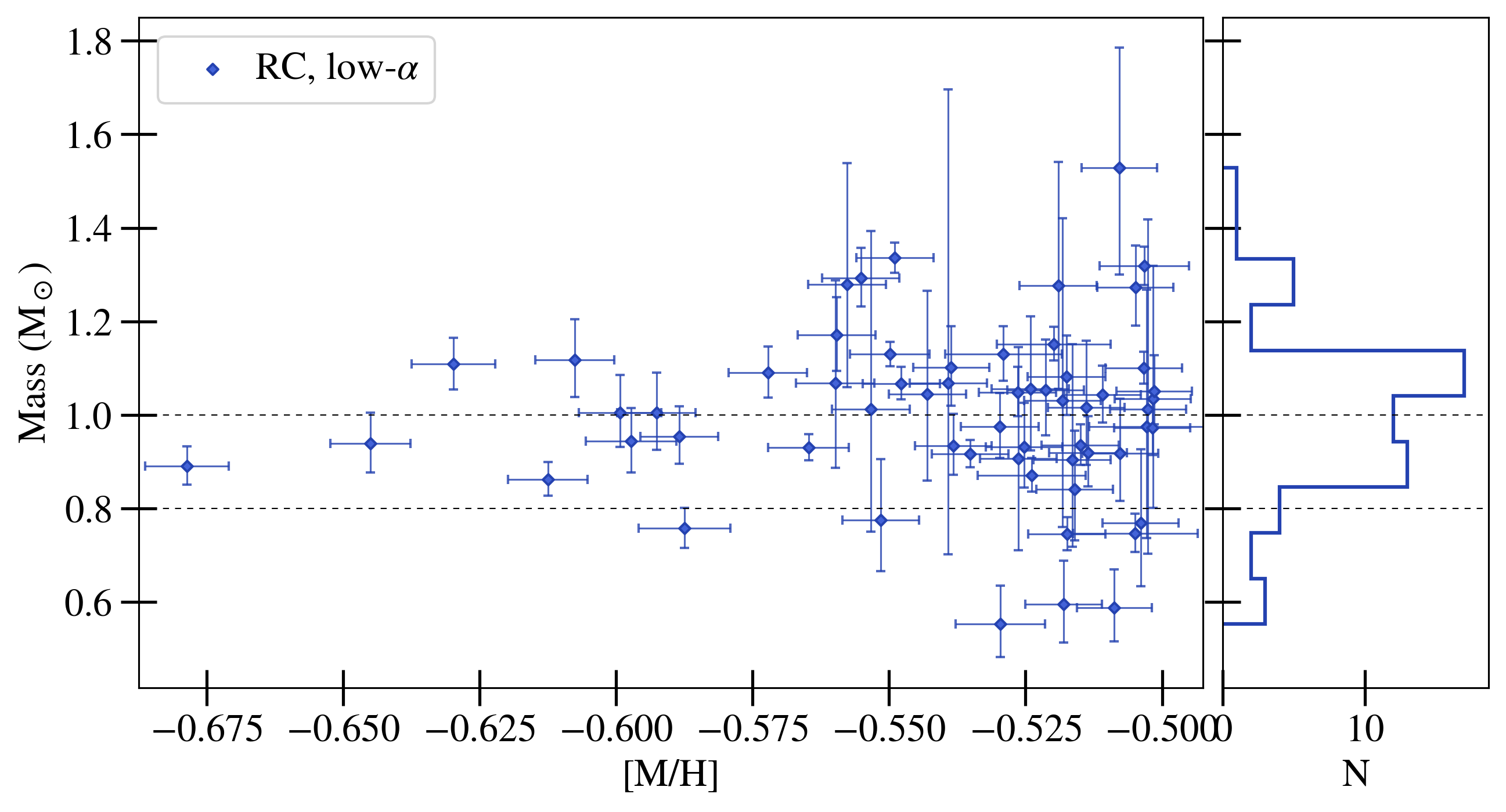}
\end{subfigure}

\begin{subfigure}
    \centering\includegraphics[scale=.4]{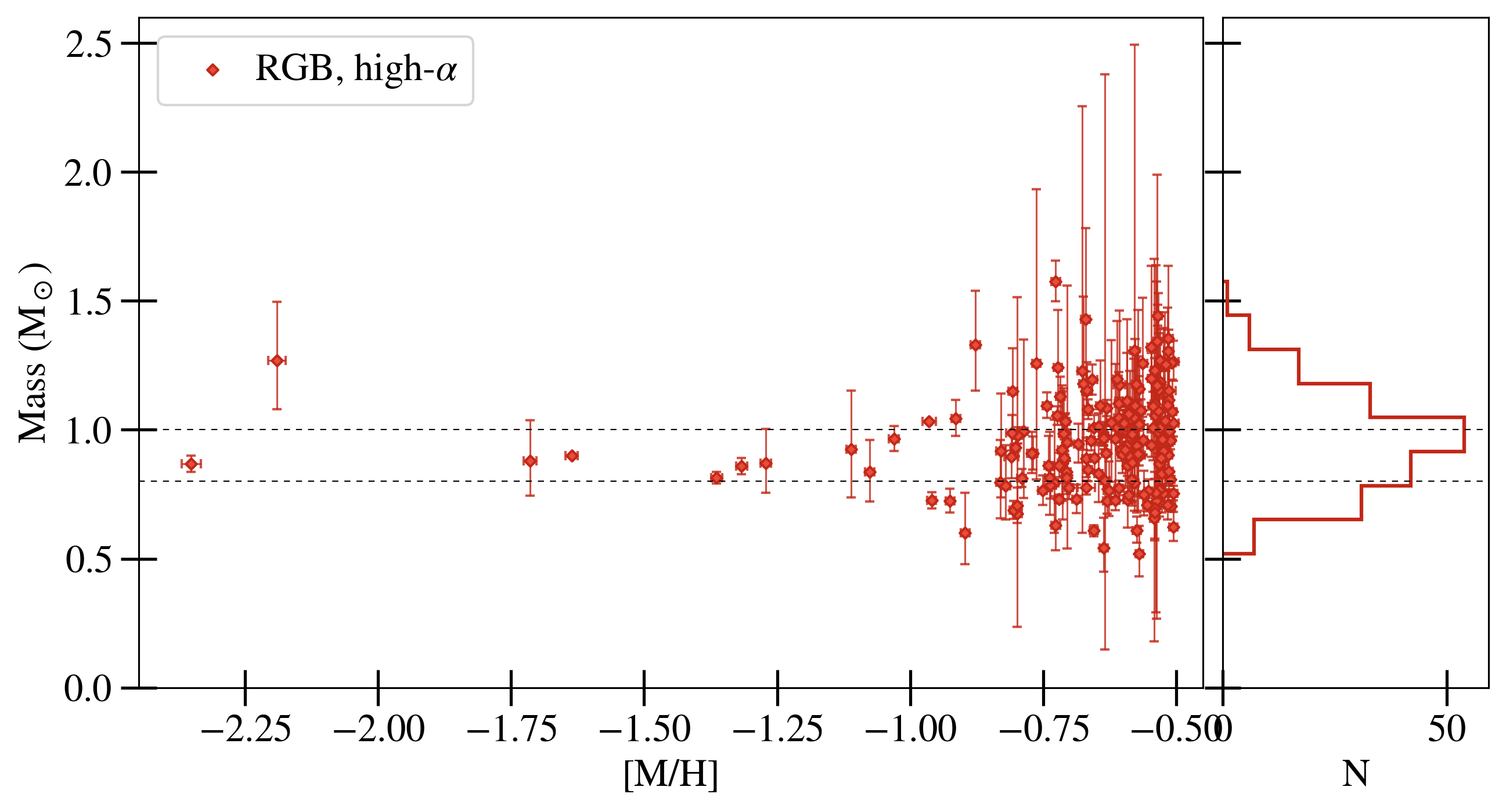}
\end{subfigure}
\begin{subfigure}
    \centering\includegraphics[scale=.4]{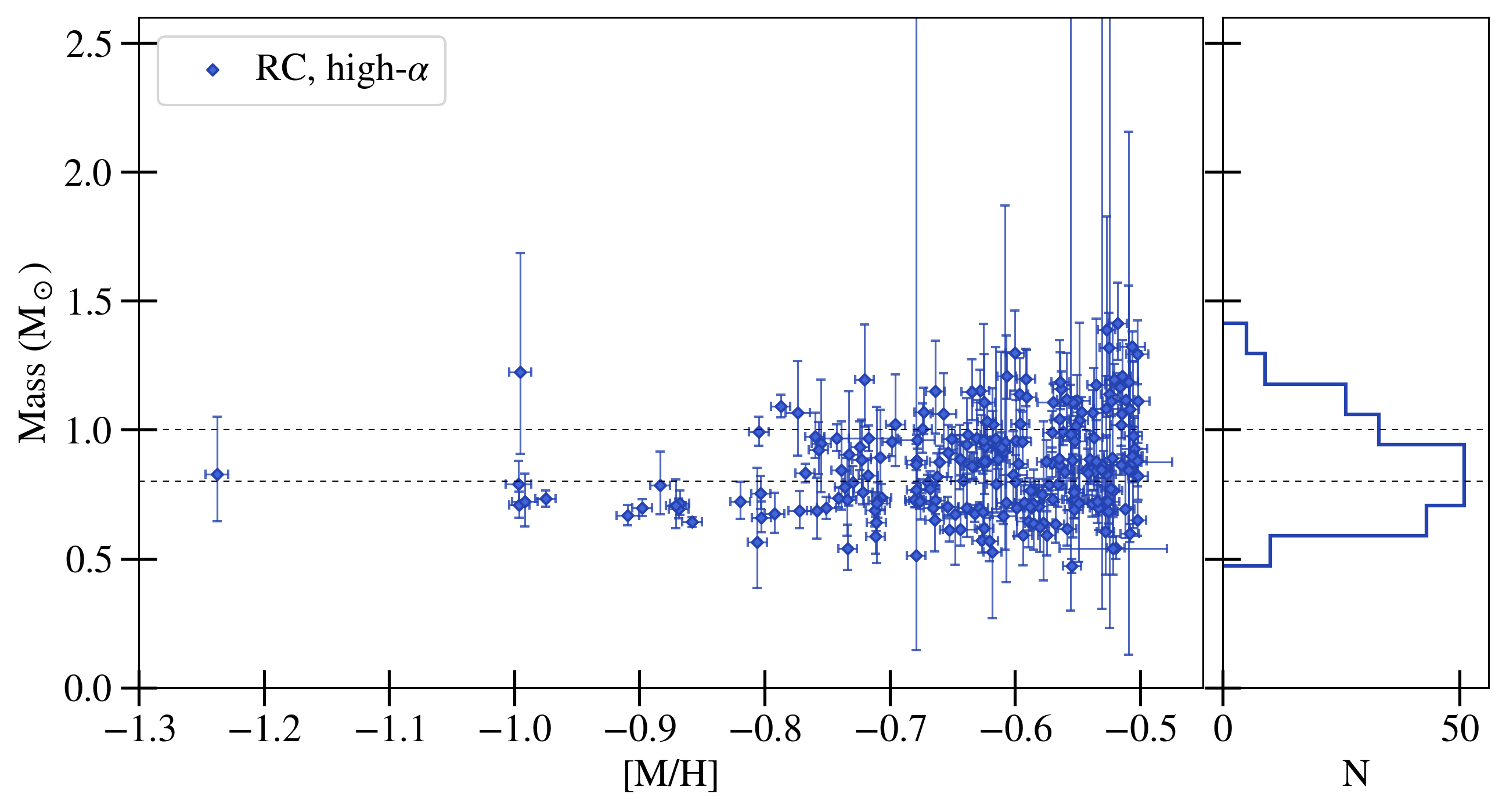}
\end{subfigure}

\caption{\label{fig:mass} The calculated masses of all stars in our sample. The RGB stars are shown on the left in red, and the RC stars are shown in blue on the right. The first row only plots the $\alpha$-poor ([$\alpha$/M] $<0.16$) stars with calculated masses, and the bottom row only plots the $\alpha$-rich ($\alpha$/M] $>0.16$) stars. The black lines show the range of masses that were expected for stars with these metallicities \citep{apok2}.}
\end{figure*}

The results of the mass calculations are shown in Figure \ref{fig:mass}. In other asteroseismic studies of low-metallicity stars, the expected range of masses for red giants in the Milky Way thick disk was approximately $0.8-1 \textrm{M}_\odot$ \citep{epstein2014,apok2}. This range is also shown in Figure \ref{fig:mass} to demonstrate how our asteroseismic masses compare to the expected mass for low-metallicity red giants.

In the past, it has been suggested that the uncorrected APOGEE temperatures are a better representation of the true temperature scale in the metal-poor regime \citep{apok2}. We found that, when calculating the mass of each star using Equation \ref{eqn:mass}, the uncorrected temperatures in APOGEE yielded lower masses than the corrected temperatures. We therefore chose to use the uncorrected temperatures, which decreased the asteroseismic mass of each star, so that the average mass of the stars fell from $0.96\textrm{M}_\odot$ to $0.93\textrm{M}_\odot$ (which is significant given that the median error is $\pm 0.08\textrm{M}_\odot$), closer to the center of the expected range from \citet{epstein2014} ($0.8-1 \textrm{M}_\odot$). Additionally, using the uncorrected temperatures resulted in $34.6\%$ of the sample falling within the expected band of $0.8$ and $1 \textrm{M}_\odot$, while the corrected temperature values yielded expected masses for $32.8\%$ of stars.


The $\alpha$-rich stars in our sample (stars with [$\alpha$/M] $>0.16$ according to APOGEE were considered $\alpha$-rich for this metallicity regime) are expected be the oldest in the set, and therefore the least massive. Their composition suggests that they formed in a cloud containing a high abundance of $\alpha$ elements---likely in the galaxy during the early universe before many of the first Type Ia supernova enriched the ISM with iron-peak elements \citep{burbidge1957,timmes1995}. Figure \ref{fig:mass} isolates the masses of stars with high $\alpha$ abundances. When plotting only high-$\alpha$ stars, the mass distribution, represented by the histogram, is skewed towards lower masses, corresponding to older stars, and is therefore closer to what we would expect for this regime of $\alpha$ abundances.

Overall, the masses are higher than expected for low-metallicity red giants, a known trend that has been observed in other studies using asteroseismic scaling relations \citep{epstein2014,apok2}. The high-$\alpha$ RGB stars should have masses closest to the expected range because their $\alpha$ abundance more strongly suggests an old age, and they have not yet undergone possible mass loss in the transition to the RC state \citep{howell2022,tailo2022}. Even when just looking at the high-$\alpha$ RGB stars, only $37.2\%$ of the sample's masses fall in the expected range of 
$0.8-1 \textrm{M}_\odot$. However, when this sample is limited to the $10$ high-$\alpha$ RGB stars with the lowest metallicities ($\textrm{[M/H]}<-1.0$), this percentage rises to $90\%$.

\subsubsection{Binary Interactions}\label{sec:binaries}

A significant portion of the $\alpha$-rich RGB stars in this set ($21.5\%$) have a mass greater than $1.1\textrm{M}_\odot$. This roughly corresponds to an age of $<6$ Gyrs using our age model (see Section \ref{sec:ages}), which would be unexpectedly young. Their chemical composition indicates that these stars must have formed in a time where supernovae had not yet polluted the ISM with heavier elements, so either they must be forming in a chemically unusual portion of the galaxy \citep{chiappini2015,anders2017} or they must have interacted with a companion that changed their mass \citep{martig2015,silva2018}.

Figure \ref{fig:astero} is an example of a high mass star ($1.443\pm0.040$ M$_\odot$) with clear oscillations and therefore a reliable $\Delta\nu$, $\nu_\textrm{max}$, and mass. Its seismic radius also matches the Gaia radius well, with a ratio of $0.99 \pm 0.05$. Additionally, the giant has a metallicity of $-0.534\pm0.007$ and a high $\alpha$-abundance ([$\alpha$/M]$=0.262\pm0.007$), so we argue that it is most likely an old star with an unreasonably high mass. This star and others in our set with atypical masses could be either mergers or have undergone mass transfer due to a stellar companion, leading to a sizeable number of stars with a mass that does not correlate with their chemistry or accurately predict their age.

The close binary fraction of solar-type stars is known to increase at low metallicities, so we believe that this high rate of atypical-mass stars in our sample reflect that phenomenon. In \citet{moe2019}, the close binary fraction is calculated based the fraction of spectroscopic binaries corrected for selection bias. At the metallicity range where we measured abnormally high masses (stars with $\textrm{M}>1.2\textrm{M}_\odot$ from approximately [M/H]$=-1$ to $-0.5$), the close binary fraction is expected to be $30-50\%$, rather than the $20\%$ expected for solar metallicity stars \citep{moe2019}.

Based on measures of stellar rotation with spectroscopy, approximately $15\%-25\%$ of solar metallicity red giants are estimated to have either tidally interacted with, merged with, or accreted material from a companion star on the RGB \citep{binaries2024}. These results closely align with \citet{moe2019}, so we similarly expect an even higher fraction of binary interactions at the lower metallicity of our sample.

This could reasonably explain the $21.5\%$ of $\alpha$-rich RGB stars in this set with large masses ($>1.1$ M$_\odot$) and the $6.3\%$ with low masses ($<0.7$ M$_\odot$). However, not all close binary systems will have exchanged more than $0.1$ M$_\odot$ of material, which explains why our total percentage of abnormal masses ($27.7\%$) is on the low end of 
the close binary fractions predicted in \citep{moe2019} for low-metallicity stars.


\subsubsection{Mass Loss}\label{sec:mass_loss}

\begin{figure}
\centering\includegraphics[scale=.38]{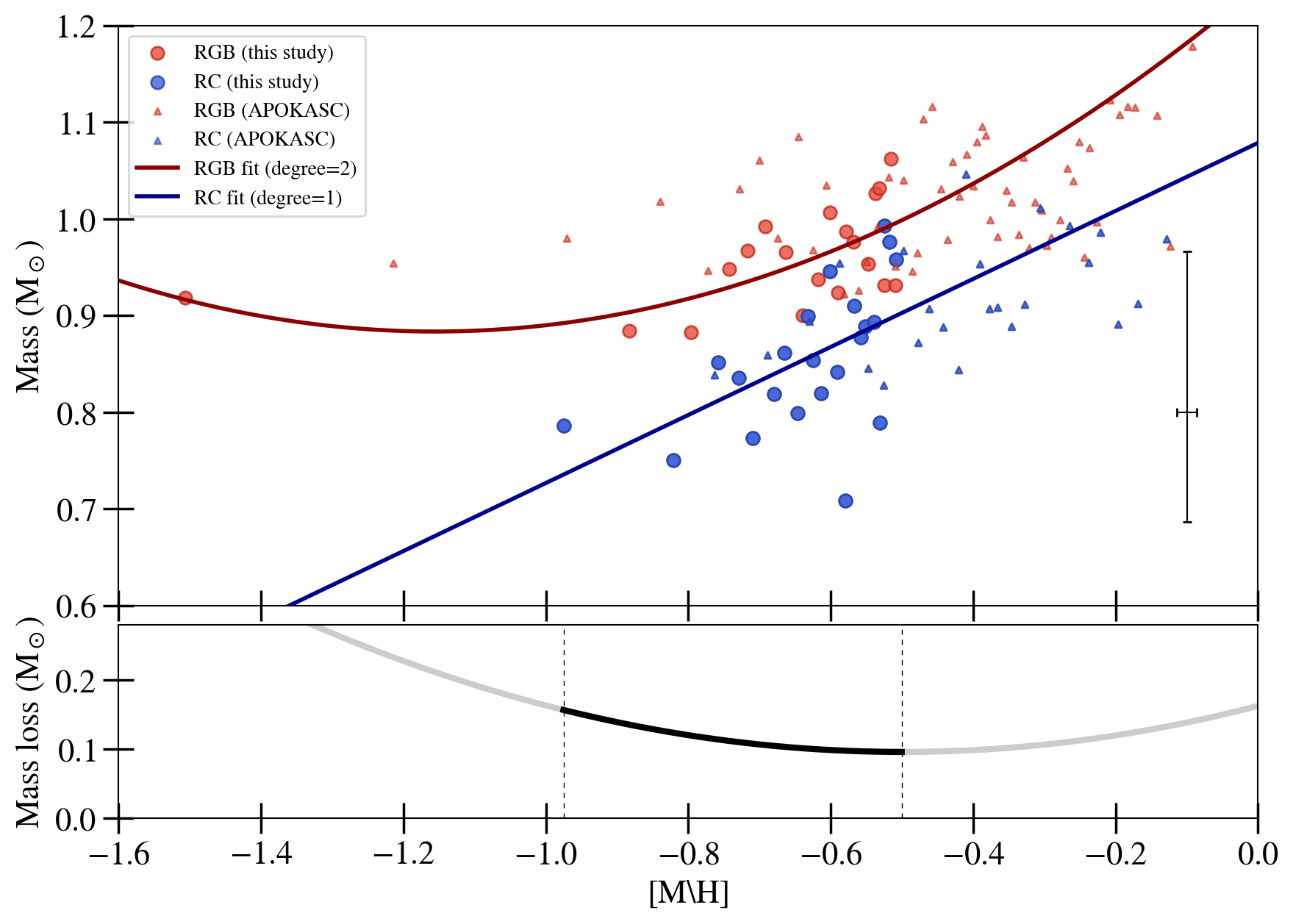}
\caption{\label{fig:mass_loss} The top figure shows two lines fitted to the high-$\alpha$ stars from this set, with a second-degree polynomial fit to the RGB stars and a first-degree polynomial to the RC stars. For clarity, the stars have been binned by metallicity into groups of $10$. The bottom plot shows the difference between the red (RGB) and blue (RC) lines to estimate mass loss. Our mass loss estimate has been limited to the section between the vertical lines, where we have calculated masses for both types of stars. The APOKASC data is shown in smaller points (also binned into groups of $10$) to demonstrate that this trend appears to continue even in stars above [M/H] $=-0.5$.}
\end{figure}

Another notable feature of Figure \ref{fig:mass} is that the distribution of RC stars is skewed towards lower masses than the RGB stars. This likely reflects the mass loss that is known to occur between the RGB and RC evolutionary stages \citep{miglio2012,tailo2022,howell2022}. This was further examined in Figure \ref{fig:mass_loss}, where lines were fit to the high-$\alpha$ stars (presumably a uniformly old population) from the lower plots in Figure \ref{fig:mass}. When the RGB and RC stars are fitted separately, both with a second-degree polynomial, the masses seem to diverge at lower metallicities. This indicates that mass loss when evolving from the RGB to RC state is metallicity-dependent, something that has been noted previously \citep{tailo2022,tayar2023}. The APOKASC data, although not included in the fit, is shown to demonstrate that the lines fit to the masses from this study appear 
to somewhat agree with the data at those higher metallicities.
\\

\subsection{Comparison to Other Studies}

 \begin{figure}
\centering\includegraphics[scale=.46]{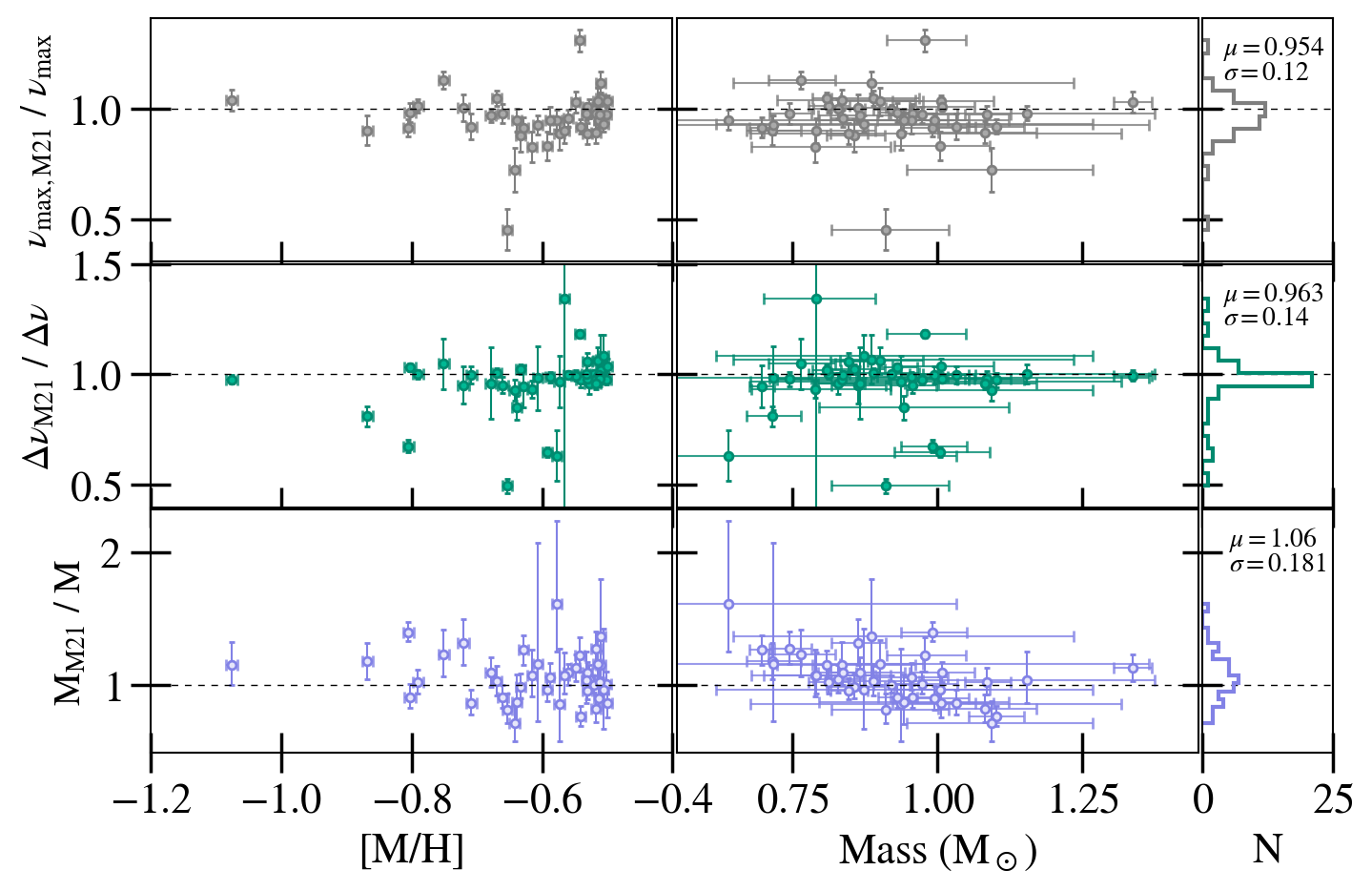}
\caption{\label{fig:ted_comp} The $\Delta\nu$, $\nu_\textrm{max}$, and corrected masses of this study compared to those calculated using the BHM pipeline \citep{bhm2020} in \citet{mackereth2021}.}
\end{figure}

Some of our stars were also studied by \citet{mackereth2021}, and we compared their results to our values when available (Figure \ref{fig:ted_comp}). The $\Delta\nu$ and $\nu_\textrm{max}$ values match well, with a large majority falling close to $1$. The ratio of masses, however, follow a trend when plotted as a function of mass. For masses that we determined to be lower than $~1$ M$_\odot$, the masses from \citet{mackereth2021} become increasingly deviant from ours. This is likely because of the careful analysis needed to calculate the masses of low-metallicity stars. For example, the corrections made using \texttt{asfgrid} are metallicity-dependent, which is necessary to consider in the low-metallicity regime, and were not applied in \citet{mackereth2021}. These corrections kept our calculated masses closer to the lower masses presumed to be associated with our low-metallicity stars \citep{epstein2014}.

\subsection{Ages}\label{sec:ages}

To estimate the ages for each of our stars, we interpolated a grid from \citet{tayar2017}, which had previously been used to compute ages for asteroseismic stars. 
This grid specifically accounts for $\alpha$-element enhancement and spans both the mass and metallicity range of our sample.


The model considered the mass, log$g$, metallicity, and [$\alpha$/M] abundance for each star to calculate an age. The asteroseismic masses (which were calculated as explained in Section \ref{sec:mass}) and asteroseismic log$g$s, (calculated similarly to mass, but instead using Equation \ref{eqn:logg}
), were considered along with the abundances and uncorrected temperatures from APOGEE. The estimated mass loss calculated in \ref{sec:mass_loss} was added to the asteroseismic masses of the red clump stars with a metallicity from $-1.2<$[M/H]$<-0.5$ before calculating an age.


\begin{figure*}
\begin{subfigure}
    \centering\includegraphics[scale=.4]{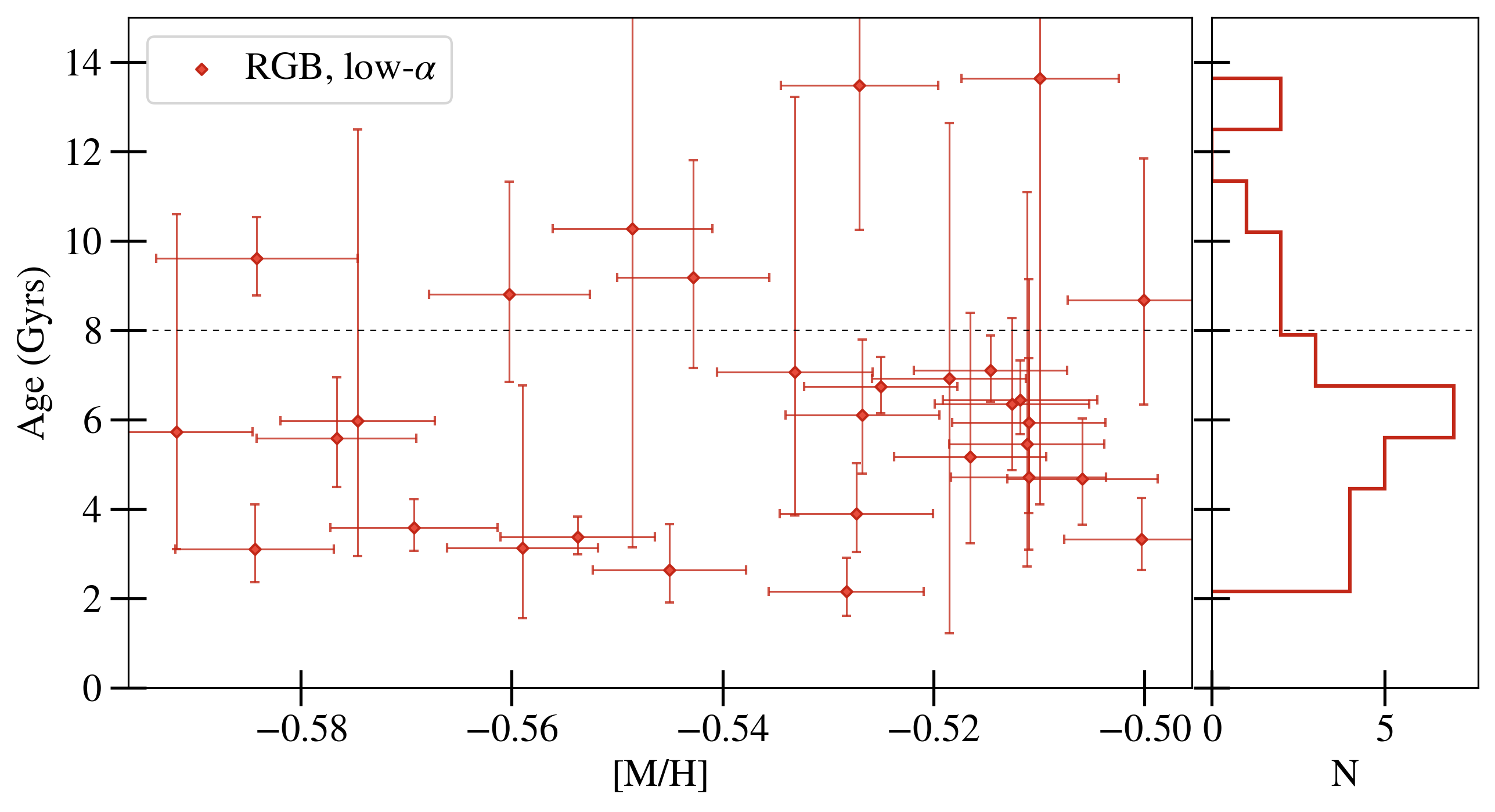}
\end{subfigure}
\begin{subfigure}
    \centering\includegraphics[scale=.4]{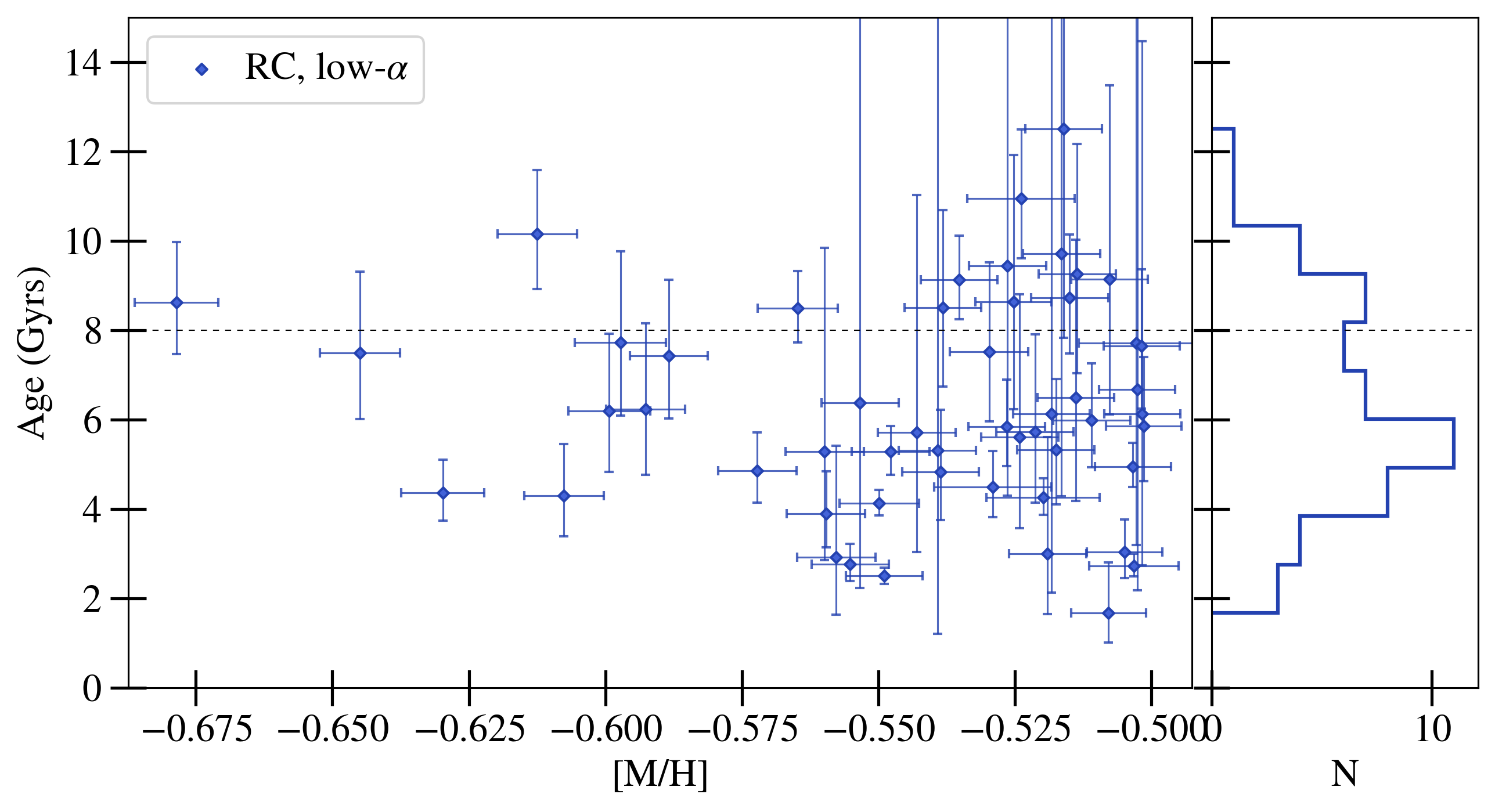}
\end{subfigure}

\begin{subfigure}
    \centering\includegraphics[scale=.4]{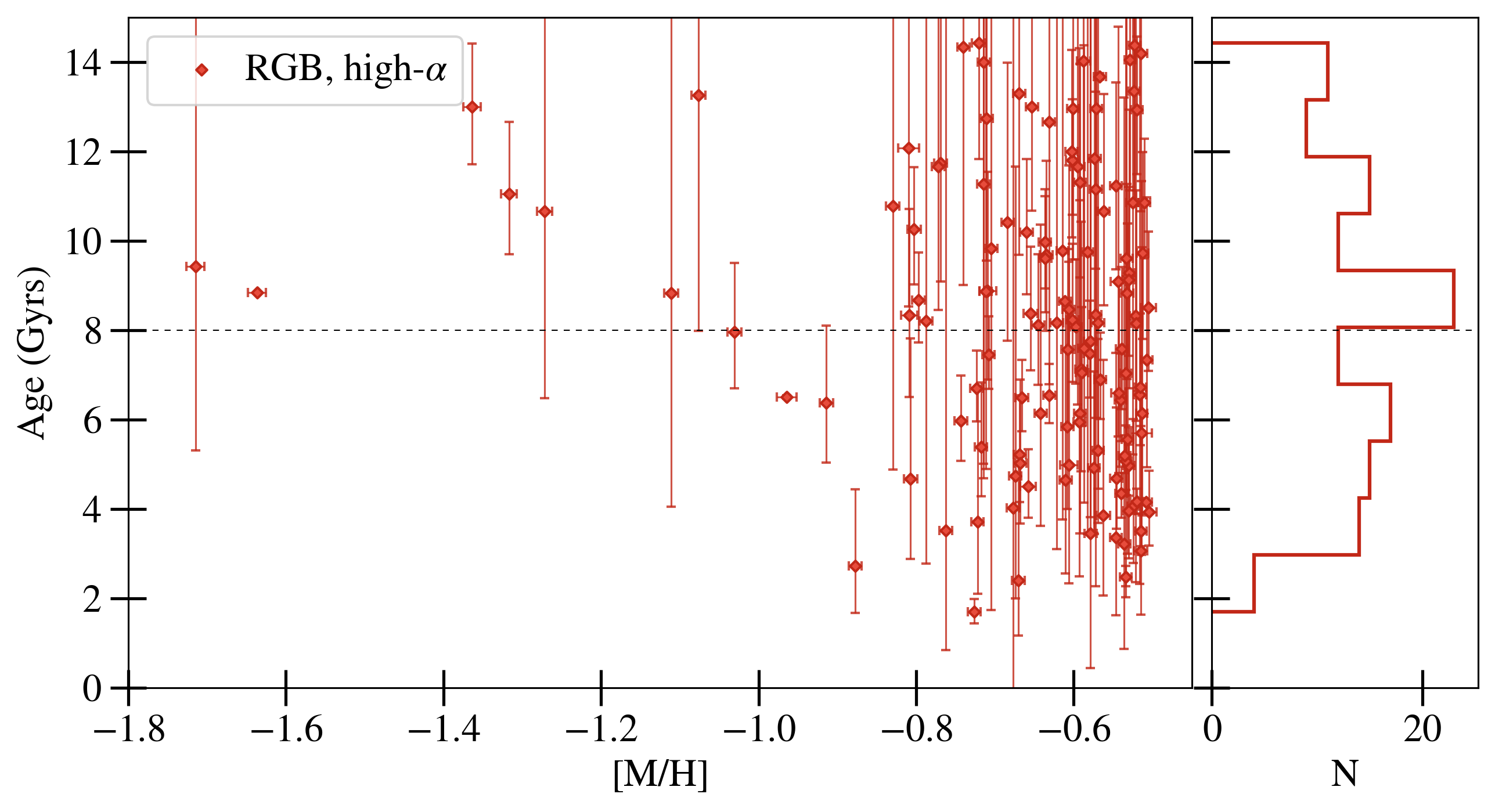}
\end{subfigure}
\begin{subfigure}
    \centering\includegraphics[scale=.4]{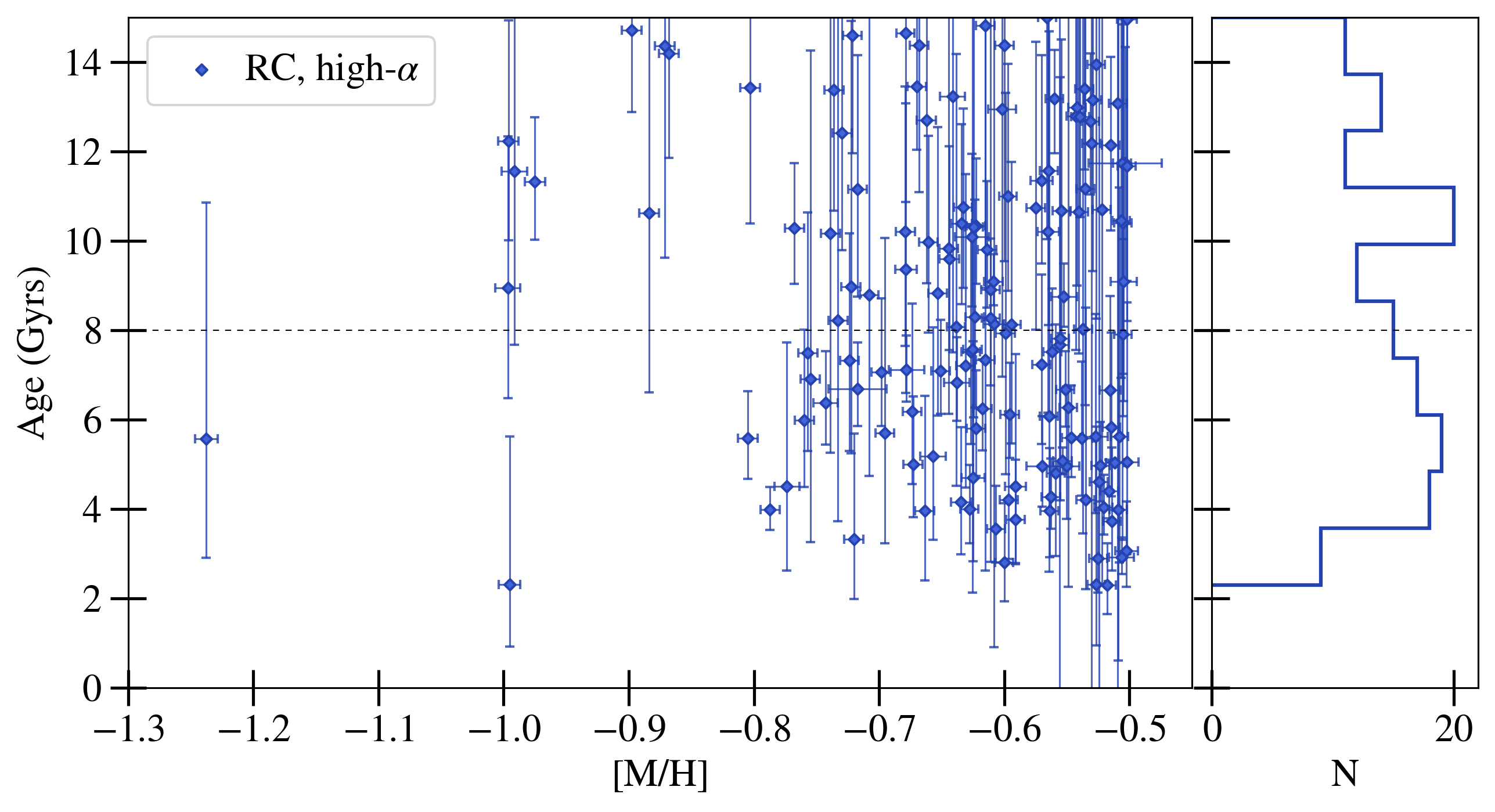}
\end{subfigure}

\caption{\label{fig:age} The derived ages of all stars in our sample, cut off at $15$ Gyrs. The RGB stars are shown on the left in red, and the RC stars are shown in blue on the right. The first row plots all of the stars with calculated ages, and the bottom row only plots the $\alpha$-rich stars. The ages are plotted in Gyrs. The dotted line at $8$ Gyrs approximately represents the age of a star with $\textrm{M}=1\textrm{M}_\odot$.}
\end{figure*}

Using this model grid, we were able to determine the ages of $506$ stars out of the $515$ with an asteroseismic mass. The results are shown in Figure \ref{fig:age} for all the stars with ages $<15$ Gyrs. It should be noted that the ages of $28\%$ of our sample were calculated as above $15$ Gyrs, which is unrealistically older than the age of the universe. We expect that these high ages are related to stars with low masses as a result of binary interactions, as discussed in Section \ref{sec:binaries} These are included in the data table and relevant calculations, but are not depicted in Figure \ref{fig:age}.

As expected, the distribution of the $\alpha$-poor stars is skewed towards younger ages than the $\alpha$-rich stars, which was expected based on Figure \ref{fig:mass}. All the figures show a significant population of stars with ages $<8$ Gyr (roughly translates to a mass of about $>1\textrm{M}_\odot$), which could be partially due to binary interactions inflating the masses of some stars out of the expected range, as discussed in Section \ref{sec:binaries}, and causing a fraction of stars appearing to have younger ages than expected in each plot of Figure \ref{fig:age}.

Focusing on the $\alpha$-rich stars in the bottom two plots of Figure \ref{fig:age}, there are noticeable increases in the number of stars at $\sim8-10$ Gyr, or close to the age of the universe. This is more obvious in the RGB stars, because, as we established in Figure \ref{fig:mass}, 
many RC stars in this sample have a significantly lower mass, leading to higher ages, even after adding the estimated mass loss calculated in Section \ref{sec:mass_loss} before the age determination.

\subsubsection{Young \texorpdfstring{$\alpha$}{a}-rich Stars}\label{sec:young_alphas}

The increase in the number of young $\alpha$-rich stars with an age of $\sim5$ Gyrs appears abnormal, 
as a higher $\alpha$ abundance typically indicates an older age (as mentioned in Section \ref{sec:mass}), but has been observed in previous studies \citep{chiappini2015,martig2015,grisoni2024}. However, the origin of this population is still unclear.

To investigate the cause of the young $\alpha$-rich stars, the kinematic and chemical properties of the high-$\alpha$ stars with ages less than $8$ Gyrs were investigated. Some have suggested that they may have a directional dependence \citep{martig2015}, but we did not observe any among these stars. Additionally, we did not find any other significant kinematic or chemical abnormalities shared by the young $\alpha$-rich stars.

This absence of a shared kinematic or chemical abnormality could indicate that the stars appear ``young" because they are binary mergers instead. These results are similar to \citet{grisoni2024}, which, like this study, observed larger parts of the galaxy than \citet{martig2015} and also found that ``young" $\alpha$-rich stars did not have a directional dependence. Additionally, they occur at similar rates to what we would expect from binaries \citep{binaries2024}, and recent studies \citep{rui2021,kuszlewicz2023} notably have found asteroseismic evidence of mass accretion in a known ``young" $\alpha$-rich star, further suggesting that they could be binary mergers.

\subsection{Abundances}\label{sec:abun}

\subsubsection{[\texorpdfstring{$\alpha$}{a}/M]}

\begin{figure}
\centering\includegraphics[scale=.58]{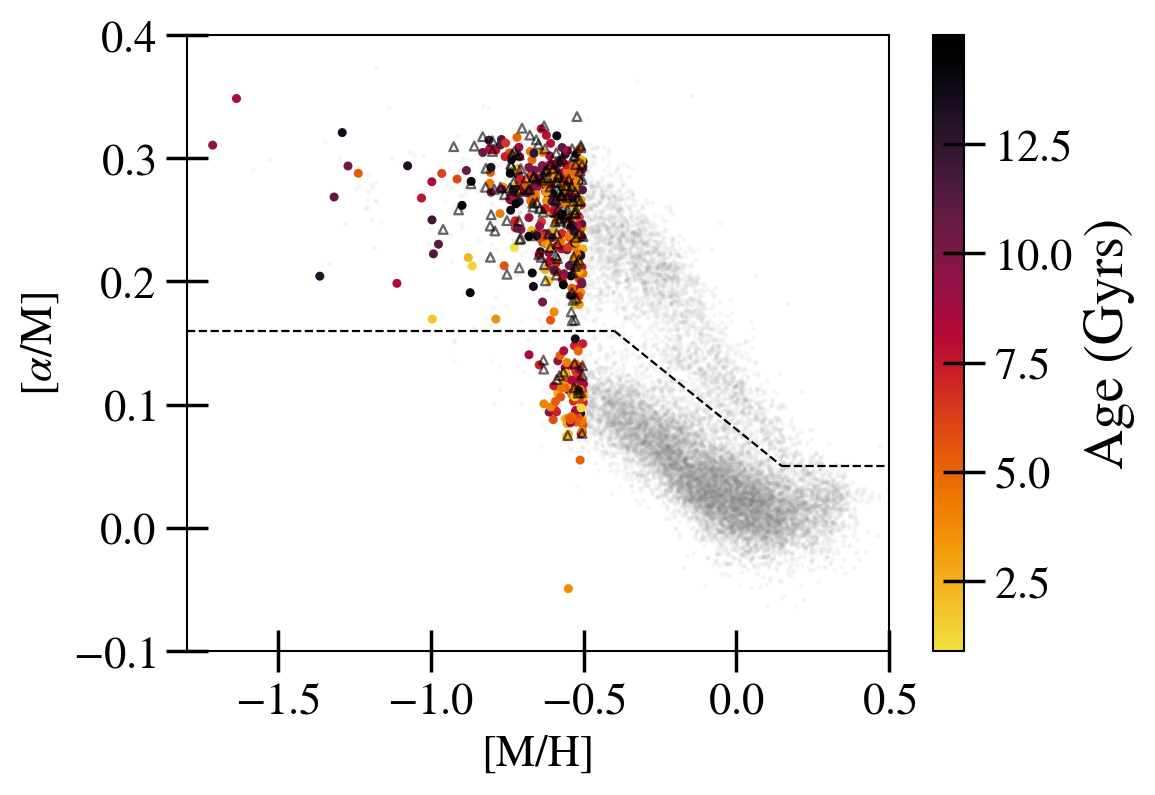}
\caption{\label{fig:abundance} The $\alpha$ abundances of the stars plotted over metallicity. The dotted black line separates the high- and low-$\alpha$ stars. Each star with an age calculated in this study has been colored according to its age. Stars with ages above $15$ Gyrs are included as the empty triangles and all the APOGEE stars are included as the smaller grey dots.}
\end{figure}

The $\alpha$ abundance ratios of our low-metallicity stars have been plotted with the other the other APOGEE DR17 stars \citep{apogee2017} in Figure \ref{fig:abundance} to demonstrate the Milky Way's $\alpha$ bimodality and its dependence on metallicity. The dotted black line shows the cutoff between high- and low-$\alpha$ stars, based on similar distinctions made in \citet{mackereth2019}, \citet{apok2}, and \citet{warfield2024}. 
This study finds that the median for stars with [$\alpha$/M] $>0.16$ is $10.8 \pm 4.3$ Gyrs and below [$\alpha$/M] $<0.16$, it is significantly lower, at $6.3 \pm 3.0$ Gyrs, which is consistent with age expectations for high- and low-$\alpha$ stars.

\subsubsection{[C/N]}

\begin{figure}
\centering\includegraphics[scale=.45]{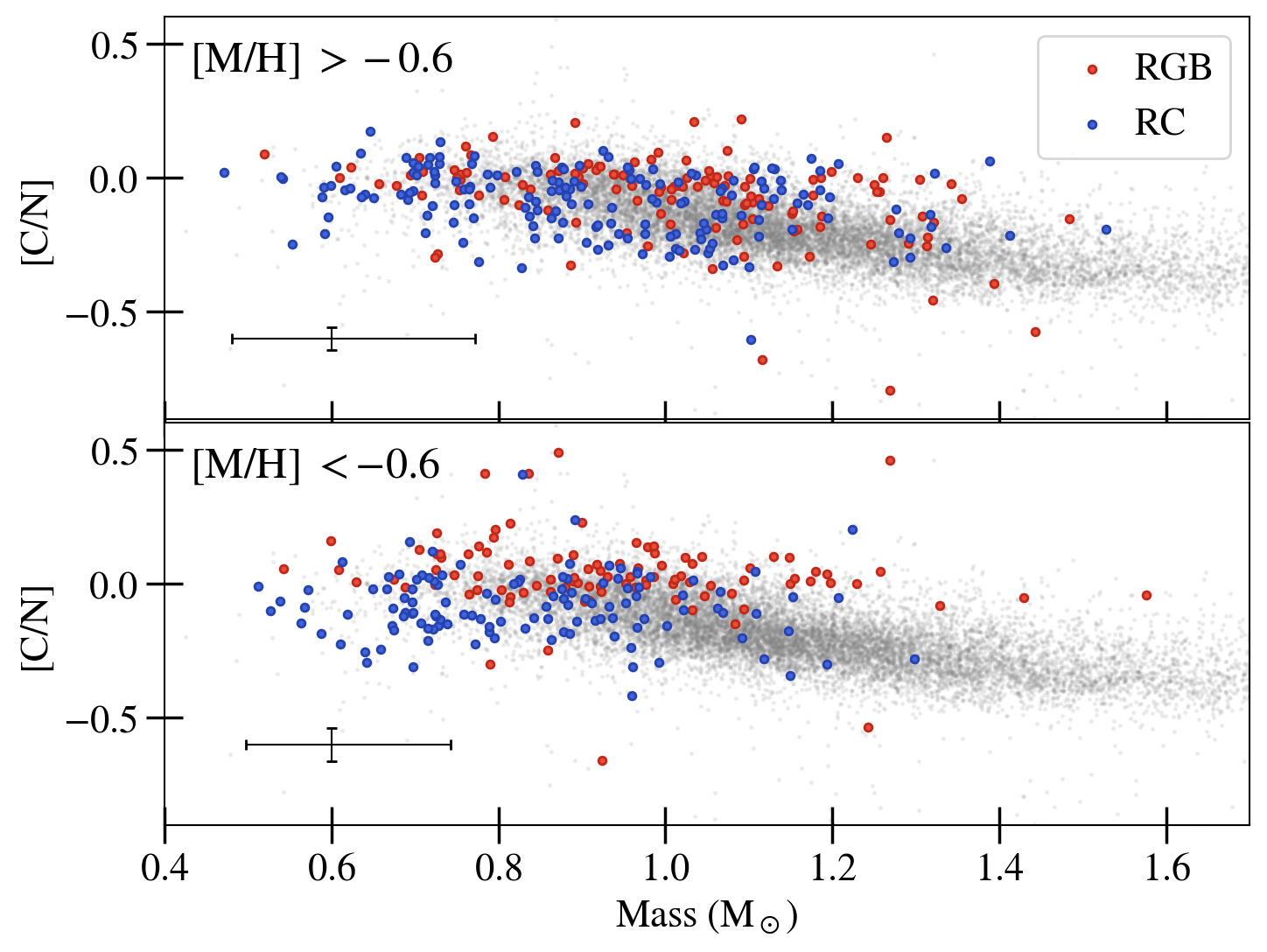}
\caption{\label{fig:c_n} The relationship between each star's [C/N] ratio and mass. The above plot shows this relationship for stars from our set with metallicities $> -0.6$. The plot below that shows this relationship for the most metal-poor stars in our set ([M/H] $< -0.6$). Each plot shows the stars split by their evolutionary state, with red representing the RGB stars and blue the RC stars. The average error bars for each set are drawn in the lower left of each plot.}
\end{figure}

Figure \ref{fig:c_n} shows the [C/N] ratio of the stars used for this study compared to the stars in the APOKASC-3 DR17 catalog \citep{apokasc3}, plotted over mass. The [C/N] ratio is indicative of the strength of a star's first dredge up, as the elemental ratios change due to the surface convective zone pulling up material that has previously been processed by nuclear burning. Therefore, there have been suggestions that this results in a relationship between [C/N] and age that can be used at low-metallicities in some cases \citep{spoo2024}. However, this relationship has generally been weaker for old, low-metallicity stars \citep{mackereth2021}, and it is expected that 
additional considerations, such as initial abundances \citep{shetrone2019,roberts2024} and extra mixing \citep{tautvai2010}, make age determination from a [C/N] ratio unlikely.

In both [C/N] plots in Figure \ref{fig:c_n}, we see evidence of this when comparing our low-metallicity red giants to the APOKASC stars. They do not follow the same [C/N]-mass relationship, instead exhibiting significant scatter. This scatter worsens when just looking at the lower metallicities, suggesting that this difference is metallicity-dependent, which matches results from previous studies \citep{shetrone2019,bufanda2023}.

Another interesting feature is that, in the [M/H]$<-0.6$ plot of Figure \ref{fig:c_n}, the RGB stars have the higher average [C/N] ratio compared to the RC stars. This difference in [C/N] between the evolutionary states is expected. As stars in the RC phase have undergone extra mixing, studies have predicted that their [C/N] ratios should be lower on average \citep{gratton2000,shetrone2019}. Additionally, the current masses of RC stars are lower than their birth masses due to mass loss. Together, these two effects shift more evolved stars downward and to the left of the trend.

The large scatter and overall weak correlation in Figure \ref{fig:c_n}, especially at lower metallicities, suggests that [C/N] is unfortunately a flawed mass and age indicator in this regime.



\begin{figure}
\centering\includegraphics[scale=.58]{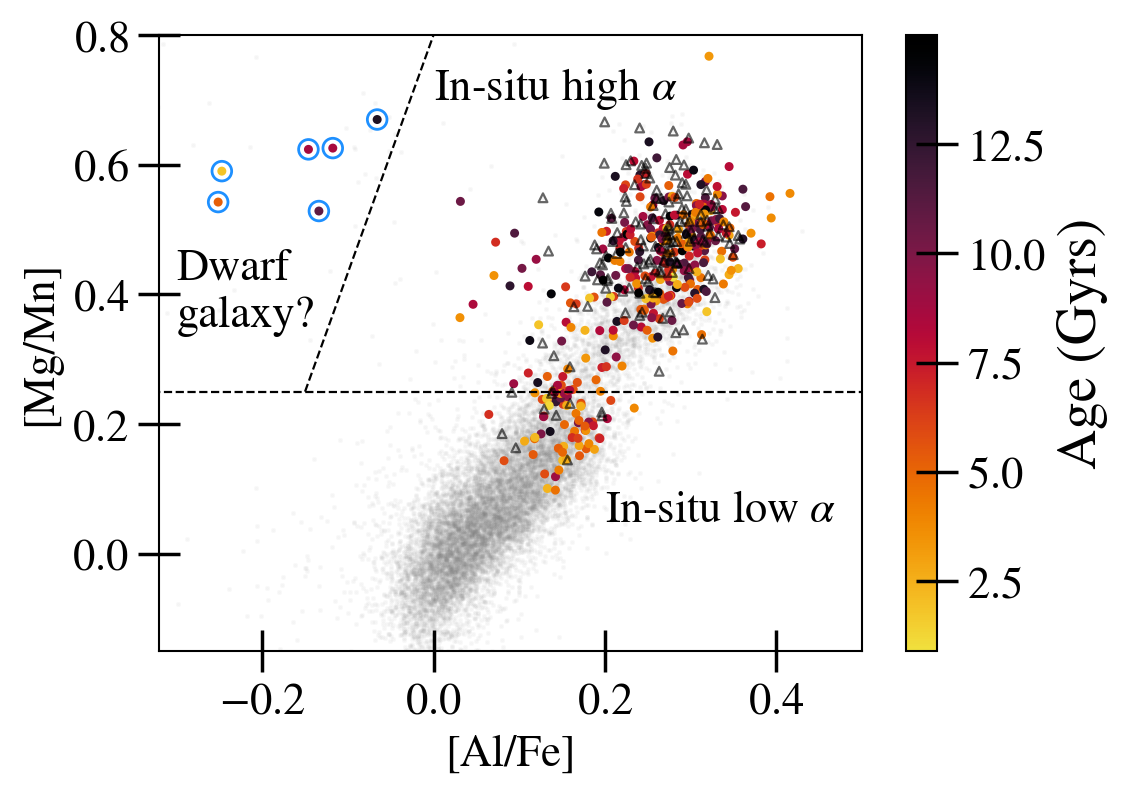}
\caption{\label{fig:mg_mn} The [Mg/Mn] ratio of each star compared to its [Al/Fe] abundance, a plot commonly used to estimate the origin of stars in the Milky Way that have migrated from nearby dwarf galaxies. This has been colored by age to display any relationships between that and the chemical abundances. Ages above $15$ Gyrs are shown as empty triangles and APOGEE stars are shown as the smaller grey dots for comparison. The lines are from \citet{dwarf_gals}. The horizontal line represents the division between high-$\alpha$ and low-$\alpha$ in-situ stars. The left of the diagonal line is where stars from dwarf galaxies that have merged with the Milky Way tend to reside. Therefore, they have been highlighted in blue and will be in the following figures as well.}
\end{figure}

\subsubsection{[Mg/Mn] and [Al/Fe]}

The relationship between [Mg/Mn] and [Al/Fe] can be used to estimate the birthplaces of stars. This is due to the fact that Mn and Mg indicate enrichment by Type Ia and Type II supernovae respectively, whereas [Al/Fe] is known to be lower in dwarf galaxies compared to in-situ stars \citep{Hasselquist2021}. Together, they can differentiate the thin and thick disk stars and isolate stars born outside of our galaxy \citep{sausage}. If we have measured ages for stars that possibly migrated from dwarf galaxies, then their ages could reveal crucial information about their birth galaxy's evolution before it merged with the Milky Way. In Figure \ref{fig:mg_mn}, there are three distinct groupings of stars. The [Mg/Mn] distribution is similar to the [$\alpha$/M] distribution; the median age of stars with a lower [Mg/Mn] ratio (below the line in Figure \ref{fig:mg_mn}) is $6.1$ Gyrs and the median of the stars above is $10.6$ Gyrs.

The majority of our stars have abundances consistent with in-situ formation, as we would expect \citep{sausage}. However, we identify six stars with a chemical composition suggesting they could have migrated from dwarf galaxies. In the next section, we will continue to highlight these stars as we compare their kinematics to the other stars in our sample.

As for the rest of the set, only stars in the Milky Way with $\textrm{[Al/Fe]}>0.0$ and $\textrm{[Mg/Mn]}>0.2$ are likely in-situ high-$\alpha$ disk stars, with some possible contamination from the Gaia Enceladus/Sausage system (GE/S) \citep{sausage}. However, as we will show in Section \ref{sec:kine}, the angular momentum (L$_\textrm{z}$) of the vast majority of the stars fall far outside the expected range of L$_\textrm{z}$s for GE/S stars, so significant contamination is unlikely.


\subsection{Kinematics}\label{sec:kine}

\begin{figure*}
\begin{subfigure}
    \centering\includegraphics[scale=0.6]{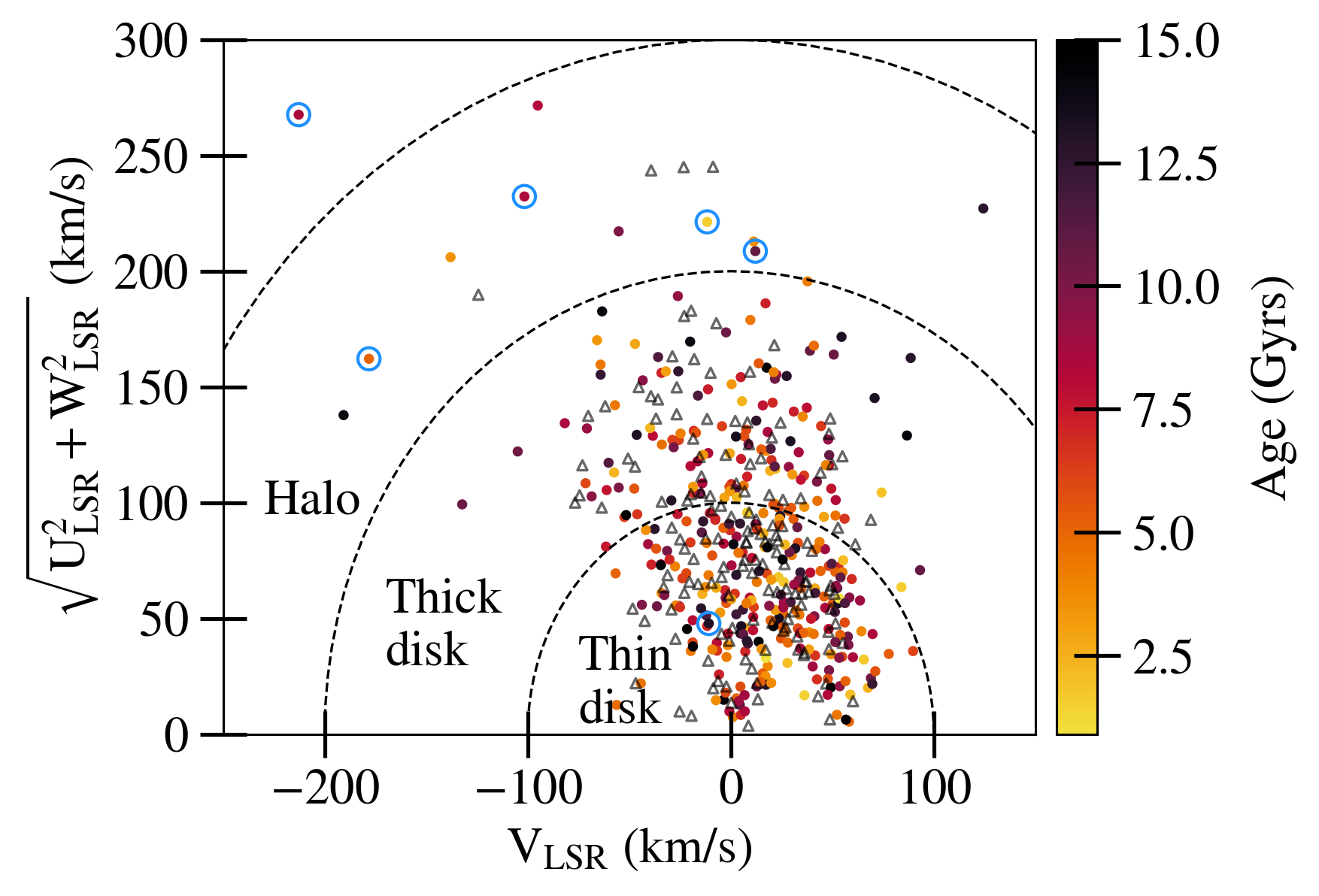}
\end{subfigure}
\begin{subfigure}
    \centering\includegraphics[scale=.65]{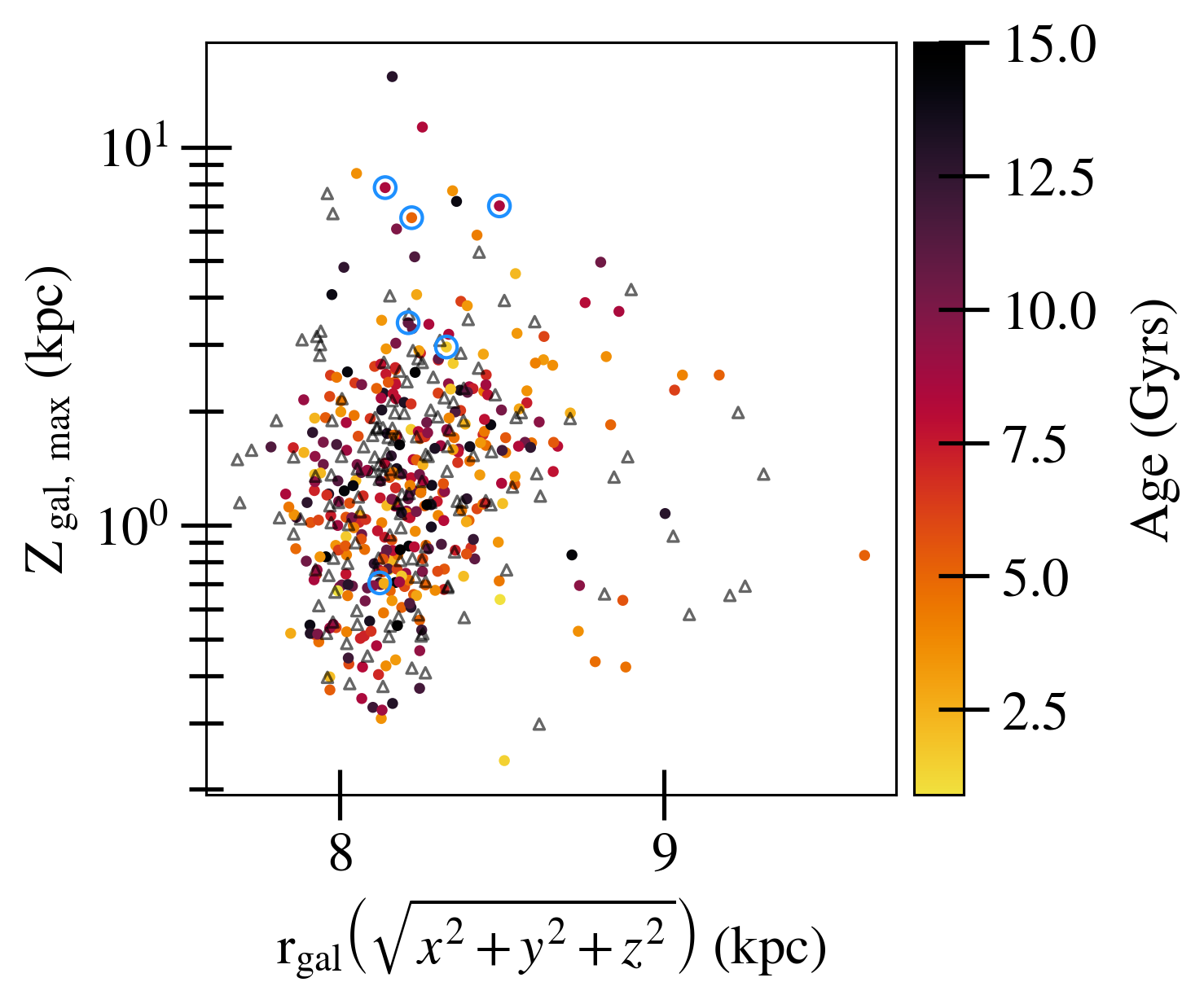}
\end{subfigure}

\begin{subfigure}
    \centering\includegraphics[scale=0.7]{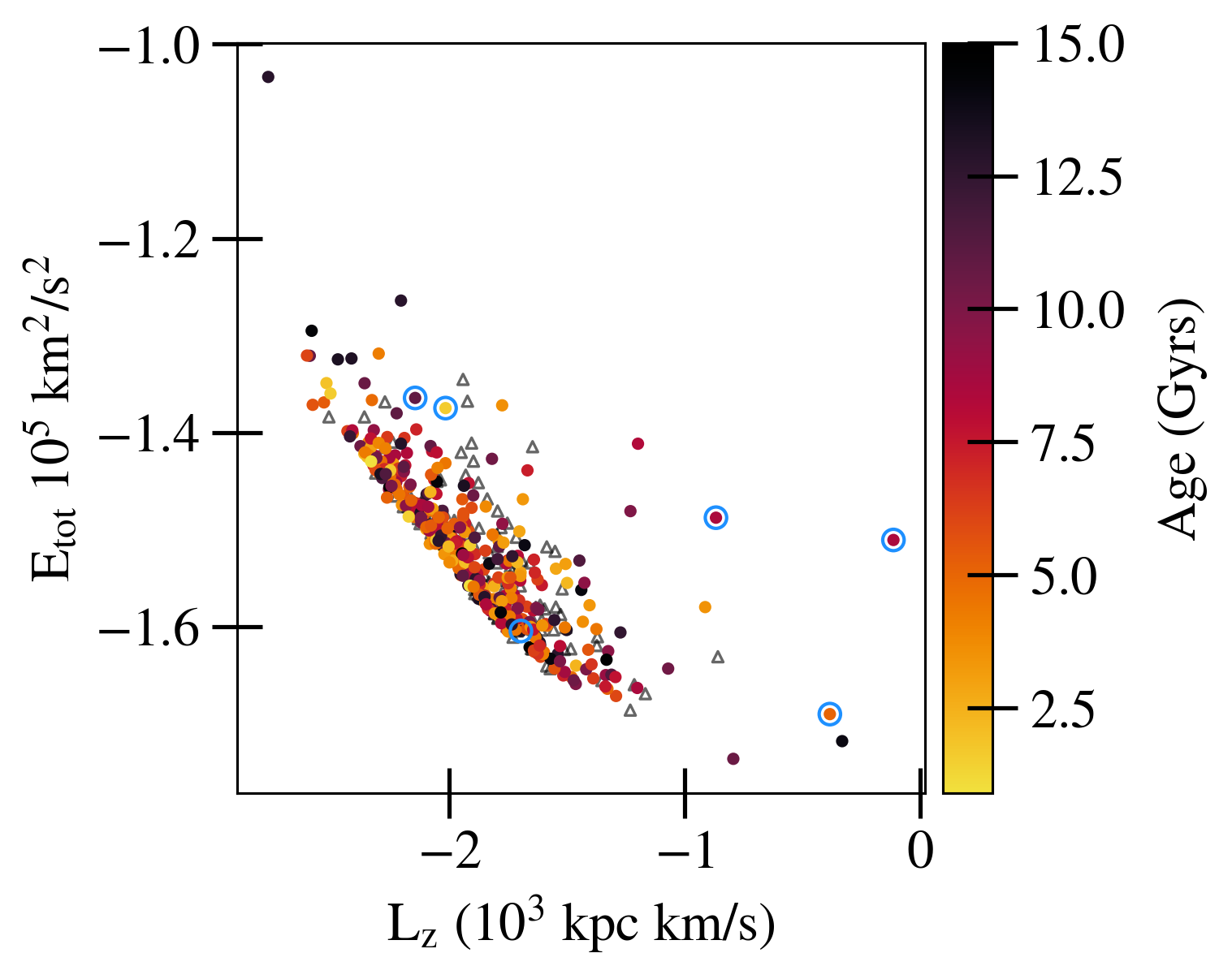}
\end{subfigure}
\begin{subfigure}
    \centering\includegraphics[scale=0.7]{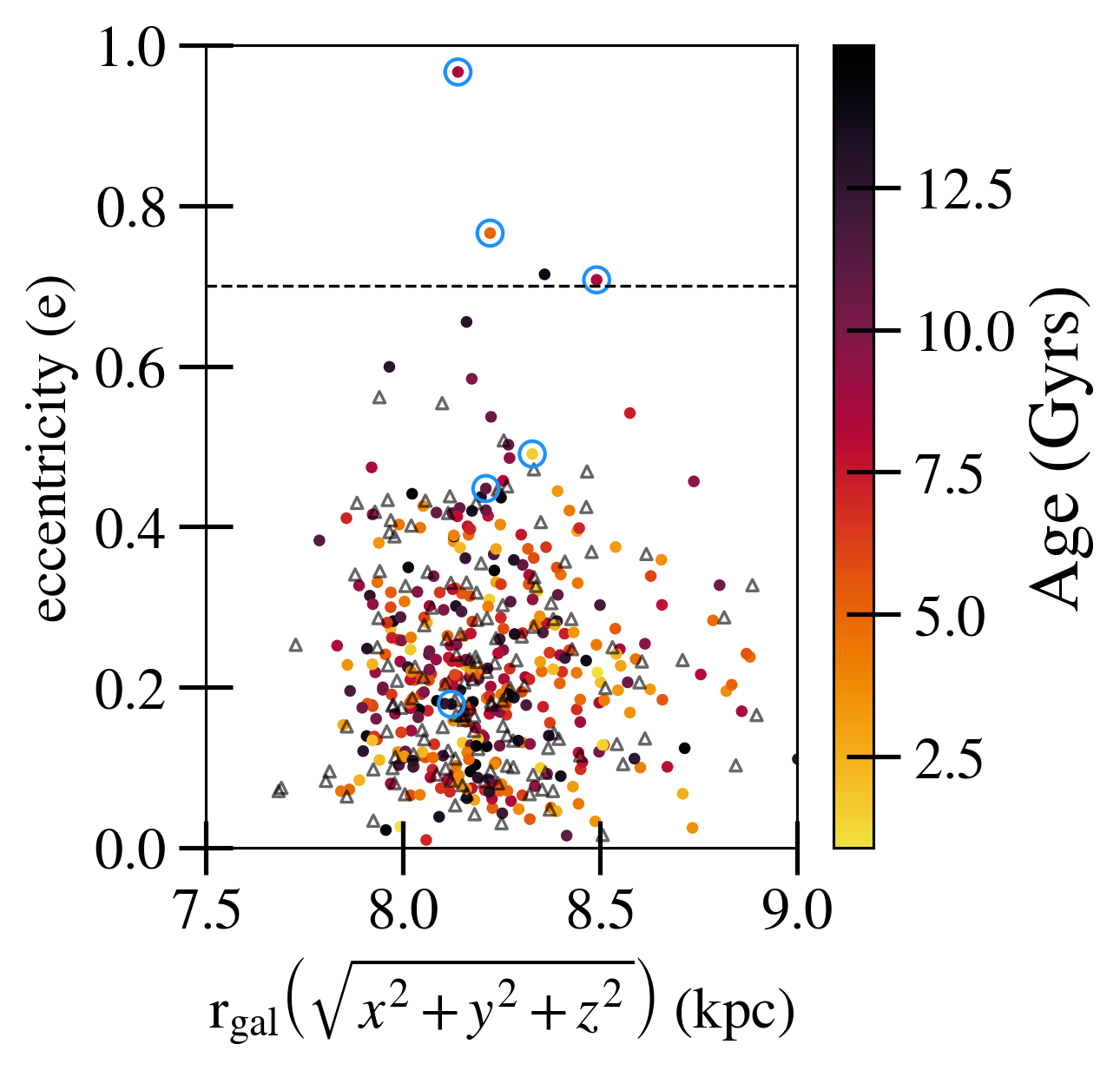}
\end{subfigure}

\caption{\label{fig:kin} The kinematics of all the stars and their respective ages. The stars with ages $>15$ Gyrs are shown as empty triangles. The top left figure shows a Toomre diagram of the stars in our set. The stars within the innermost ring are likely in the thin disk of the galaxy. Similarly, the stars between the first and second ring are probably in the thick disk, and those between the second and third ring are in the halo. The top right plot shows the maximum distance from the galactic plane in a star's orbit compared to its distance from the galactic center. The bottom left plot shows the total orbital energy and the $z$ component of its angular momentum (L$_\textrm{z}$). The bottom right plot shows eccentricity and Galactic distance. The horizontal line is from \citet{naidu2020} and is used to define the GE/S stars, which are typically selected as stars with $e>0.7$.}
\end{figure*}

In examining the kinematic properties of these stars, we can confirm what we inferred about their origin in Figure \ref{fig:mg_mn}---determining if they were born as in-situ thin or thick disk stars, halo stars, or extragalactic stars. By identifying stars with both kinematic and chemical evidence of where they were born, one could use our ages for galactic archeology to tell when those parts of the galaxy were formed. In Figure \ref{fig:kin}, different kinematic properties of the stars are depicted, with the outliers from Figure \ref{fig:mg_mn} highlighted, and the stars colored by age. These kinematics were derived using the positions and proper motions from Gaia DR3 and the radial velocities from APOGEE. The kinematic information was used to integrate the orbit of every star using \texttt{galpy} \citep{galpy}, giving us the results shown in Figure \ref{fig:kin}.

A Toomre diagram (top left of Figure \ref{fig:kin}) was made of the stars to determine the disk positions of our sample and its correlation with age. The orbital velocities of the stars suggest that most of our sample is spread throughout the thin and thick disk, with $15$ stars possibly in the galactic halo (outside the second dotted circle) \citep{venn2004}. Stars in the innermost circle of the Toomre diagram are estimated to be in-situ thin disk stars, and they have a median age of $9.7$ Gyrs. The in-situ thick disk stars are in the next circle outward, and their median age is also $9.7$ Gyrs. The halo stars (including all the stars outside the thin and thick disk) have an older median age, as expected, at $10.2$ Gyrs. Due to upside-down formation, one would expect the ages to increase further from the galactic disk \citep{bird2013}, but the top two plots of Figure \ref{fig:kin} do not show a correlation between the two. That said, we may not be able to see a strong trend given our limited sample size as well as the limited range probed in the $Z$ direction.

Interestingly, the Toomre diagram shows that five out of the six stars identified as outliers in Figure \ref{fig:mg_mn} are in the Milky Way's halo, which is consistent with them being accreted from a dwarf galaxy \citep{sausage}. This idea is further explored in the bottom two plots of Figure \ref{fig:kin}, which can be used to identify the structure that the stars originated from \citep{naidu2020}. Three of the six stars that were outliers in Figure \ref{fig:mg_mn} stand out in both plots, further indicating that they were likely not born in the Milky Way. \citet{naidu2020} defines an $e=0.7$ line that roughly divides the in-situ stars from the GE/S stars, which all have high eccentricities, leading us to believe that those three highlighted stars with $e>0.7$ could be from the GE/S system. A fourth star could be a GE/S star as well. Chemically, it was the closest star from the in-situ section of Figure \ref{fig:mg_mn} to the dwarf galaxy section, and kinematically, it has an L$_\textrm{z}$ and E$_\textrm{tot}$ comparable to our three highlighted kinematic outliers and is one of just four stars with an eccentricity above $0.7$. These stars and their ages, from left to right on the eccentricity plot in Figure \ref{fig:kin}, are: TIC 393961551 ($8.8$ Gyr), TIC 310380331 ($5.6$ Gyr), TIC 219036800 ($14.0$ Gyr), and TIC 232983181 ($8.8$ Gyr).




\begin{table*}[t]
\caption{\label{tab:table1} List of $515$ Low-Metallicity TESS Red Giants Used in this Study}
\begin{tabular}{|p{0.25\textwidth}|p{0.7\textwidth}|}
\hline
Label & Description
\\ \colrule
\hline
TIC & TIC ID\\
\hline
APOGEE & APOGEE ID\\
\hline
GAIA & Gaia ID\\
\hline
Evstate & Evolutionary state (0 for RGB, 1 for RC)\\
\hline
Sectors & Number of TESS sectors used for asteroseismology\\
\hline
dnu & $\Delta\nu$ ($\mu$Hz)\\
\hline
e\_dnu & \texttt{pySYD} error on $\Delta\nu$ ($\mu$Hz)\\
\hline
fdnu & $f_{\Delta \nu}$ from \texttt{asfgrid}\\
\hline
numax & $\nu_\textrm{max}$ ($\mu$Hz)\\
\hline
e\_numax & \texttt{pySYD} error on $\nu_\textrm{max}$ ($\mu$Hz)\\
\hline
numaxHon & $\nu_\textrm{max}$ in \citet{hon2021} ($\mu$Hz)\\
\hline
Teff\_Spec & Spectroscopic (Uncorrected) T$_\mathrm{eff}$\\
\hline
MCor & Seismic mass with f$_{\nu_\textrm{max}}$ corrections (M$_\odot$)\\
\hline
E\_MCor, e\_MCor & Upper and lower uncertainties for corrected mass (M$_\odot$)\\
\hline
Age & Asteroseismic age (Gyr)\\
\hline
E\_age, e\_age & Upper and lower error bars for ages (Gyr)\\
\hline
Rad & Gaia radius (R$_\odot$) from \citet{hon2021}\\
\hline
RadCor & Seismic radius with f$_{\nu_\textrm{max}}$ corrections (R$_\odot$)\\
\hline
E\_RadCor, e\_RadCor & Upper and lower corrected radius error bars (R$_\odot$)\\
\hline
logg & Corrected spectroscopic log$g$\\
\hline
loggCor & Seismic log$g$ with f$_{\nu_\textrm{max}}$ corrections\\
\hline
E\_loggCor, e\_loggCor & Upper and lower corrected log$g$ error bars\\
\hline
RAdeg & Right Ascension\\
\hline
DEdeg & Declination\\
\hline
plx & Parallax (arcsec)\\
\hline
Dis & Distance (pc)\\
\hline
pmRA, pmDE & Proper motion (mas yr$^{-1}$)\\
\hline
X, Y, Z & Position in X, Y, and Z direction (kpc)\\
\hline
r0 & Distance from galactic center (kpc)\\
\hline
zmax & Maximum vertical height in orbit (kpc)\\
\hline
e & Eccentricity\\
\hline
E & Orbital energy (km s$^{-1}$)$^2$\\
\hline
Lz & Z component of the angular momentum (kpc km s$^{-1}$)\\
\hline
U\_LSR, V\_LSR, W\_LSR & Heliocentric Galactic X,Y, and Z velocity (km/s)\\
\hline
M\_H & Metallicity [M/H] (dex)\\
\hline
e\_M\_H & Metallicity error (dex)\\
\hline
a\_M & [$\alpha$/M] (dex)\\
\hline
C\_N & [C/N] (dex)\\
\hline
Mg\_Mn & [Mg/Mn] (dex)\\
\hline
\end{tabular}
\tablecomments{Catalog of the calculated asteroseismology parameters and asteroseismic log$g$s, radii, masses, and ages, along with the chemical and kinematic information calculated for Sections \ref{sec:abun} and \ref{sec:kine}.}
\end{table*}

\section{Conclusion}

Through this study, we have calculated the masses and ages of $506$ low-metallicity ([M/H] $< -0.5$) red giants using asteroseismic analysis on observations from TESS. Most of these stars lie in or near the TESS CVZs and all have at least $5$ available sectors of data, covering a total field of view that is roughly $12$ times that of Kepler.

Low-metallicity stars are generally expected to be old and therefore low-mass ($0.8-1 \textrm{M}_\odot$) \citep{apok2}. We found that $34.6\%$ of our calculated masses are within that expected range, and $32.0\%$ have ages consistent with this picture ($< 15$ Gyrs and $> 8$ Gyrs). We suggest that the $21.3\%$ of red giants with masses above $1.1\textrm{M}_\odot$ are likely mergers or gained mass that was transferred from a companion star. Similarly, the $5.3\%$ of RGB stars with masses lower than $0.7\textrm{M}_\odot$ presumably lost mass during binary interactions. The greater fraction of RC stars with low masses ($18.4\%$) seem to result from mass loss in more evolved stars that we found increases at lower metallicities.

When looking at the $\alpha$-rich and RGB subsample of these stars, which are expected to be old and have more accurate asteroseismic parameter calculations, more stars have masses and ages within expectations for metal-poor red giants ($37.2\%$ of the masses and $55.0\%$ of the ages) \citep{zinn2022}. That said, we do detect that $21.5\%$ of stars in the subsample are overmassive ($>1.1 \textrm{M}_\odot$), and therefore ``young," $\alpha$-rich stars. This is slightly higher than other studies that have also identified $\alpha$-rich stars with unusually young ages \citep{martig2015,chiappini2015,grisoni2023,jofre2023}, but that is likely due to the higher binary fraction of stars in the low-metallicity regime \citep{moe2019}. Physically, we find that the ``young" $\alpha$-rich stars do not seem confined to one region in the galaxy, and thus we tend to favor a binary explanation for this subsample, although we cannot rule out a special birthplace.



For our full sample, the mass and age estimates were combined with known chemical and kinematic data to determine where our population of stars fits into the galaxy both chemically and physically. This opens the door for galactic archaeologists to compare the data, now including more accurate ages for low-metallicity stars, to galactic models.

A star's [$\alpha$/M] abundance is expected to be a rough age indicator, and we found that the average high-$\alpha$ star is indeed older than the average low-$\alpha$ star's age \citep{apok2}. We also note that the correlation between [C/N] and age is quite weak in this regime, as a combination of metallicity-dependent birth abundances, extra mixing, binary interactions, and mass loss impact their relationship \citep{shetrone2019,mackereth2021,roberts2024}.

We determined that most of the stars lie in the thin and thick disk but identify $15$ halo stars (with higher average ages than the disk stars). The expected correlation between lower masses and further distances from the galactic plane is very weak in our observations. However, this could be due to our small sample size, which limits us to a portion of the Milky Way that is too small to notice these wide-scale trends.

Six stars were outside the typical range for in-situ stars when plotting [Mg/Mn] and [Al/Fe]. Five out of the six were in the halo and had a high Z$_\textrm{max}$. Three of them, along with an additional fourth star, stood out when plotting angular momentum (L$_\textrm{z}$) and eccentricity, suggesting that they could belong to the Gaia Enceladus/Sausage substructure of the Milky Way.

This is the first study to carefully determine and verify asteroseismic masses and ages for a large sample of low-metallicity TESS stars. While we see many of the expected trends in a global sense, there seem to be interesting subpopulations that could be worthy of future study, such as the high mass stars that could be binary merger candidates and the possible GE/S stars.

By focusing on low-metallicity stars, we shine a light on a formative era of the Milky Way. We believe that these results encourage further asteroseismic study of metal-poor red giants from TESS. As TESS continues to collect data, hundreds more low-metallicity stars each year will meet the threshold of $5$ observing sectors that were needed to compute the asteroseismic masses and ages of the metal-poor stars in this study. These results, when compared to galactic evolution models \citep{andrews2017,johnson2021,lian2020}, would help to advance galactic archeology as we further explore the galaxy at various stages of evolution.

\vspace{5mm} 

The authors would like to thank Artemis Theodoridis for her contribution in cross-matching stars from APOGEE and \citet{hon2021} as well as Zachary Claytor, Christopher Lam, Leslie Morales, and Mavis Marasco
for helpful conversations that improved the quality of this manuscript.

The authors gratefully acknowledge support from the National Aeronautics and Space Administration (80NSSC23K0436).

Funding for the Sloan Digital Sky Survey IV has been provided by the Alfred P. Sloan Foundation, the U.S. Department of Energy Office of Science, and the Participating Institutions.

SDSS-IV acknowledges support and resources from the Center for High Performance Computing at the University of Utah. The SDSS website is www.sdss4.org.

SDSS-IV is managed by the Astrophysical Research Consortium for the Participating Institutions of the SDSS Collaboration including the Brazilian Participation Group, the Carnegie Institution for Science, Carnegie Mellon University, Center for Astrophysics | Harvard \& Smithsonian, the Chilean Participation Group, the French Participation Group, Instituto de Astrof\'isica de Canarias, The Johns Hopkins University, Kavli Institute for the Physics and Mathematics of the Universe (IPMU) / University of Tokyo, the Korean Participation Group, Lawrence Berkeley National Laboratory, Leibniz Institut f\"ur Astrophysik Potsdam (AIP),  Max-Planck-Institut f\"ur Astronomie (MPIA Heidelberg), Max-Planck-Institut f\"ur Astrophysik (MPA Garching), Max-Planck-Institut f\"ur Extraterrestrische Physik (MPE), National Astronomical Observatories of China, New Mexico State University, New York University, University of Notre Dame, Observat\'ario Nacional / MCTI, The Ohio State University, Pennsylvania State University, Shanghai Astronomical Observatory, United Kingdom Participation Group, Universidad Nacional Aut\'onoma de M\'exico, University of Arizona, University of Colorado Boulder, University of Oxford, University of Portsmouth, University of Utah, University of Virginia, University of Washington, University of Wisconsin, Vanderbilt University, and Yale University.

Transiting Exoplanet Survey Satellite (TESS) was accessed from \url{https://registry.opendata.aws/tess}. This paper includes data collected with the TESS mission, obtained from the AWS Open Data copy of the MAST Archive at the Space Telescope Science Institute (STScI). Funding for the TESS mission is provided by the NASA Explorer Program. STScI is operated by the Association of Universities for Research in Astronomy, Inc., under NASA contract NAS 5–26555.

This work has made use of data from the European Space Agency (ESA) mission {\it Gaia} (\url{https://www.cosmos.esa.int/gaia}), processed by the {\it Gaia} Data Processing and Analysis Consortium (DPAC, \url{https://www.cosmos.esa.int/web/gaia/dpac/consortium}). Funding for the DPAC has been provided by national institutions, in particular the institutions participating in the {\it Gaia} Multilateral Agreement.

\software{\texttt{pySYD} \citep{pysyd}, 
\texttt{Lightkurve} \citep{lightkurve}, 
\texttt{Astropy} \citep{astropy2013,astropy2018,astropy2022}, 
\texttt{galpy} \citep{galpy}, 
\texttt{echelle} \citep{echelle}, 
\texttt{ASFGrid} \citep{asfgrid,asfgrid6}, 
\texttt{NumPy} \citep{numpy}, 
\texttt{SciPy} \citep{scipy}, 
\texttt{Matplotlib} \citep{matplotlib}, 
\texttt{Pandas} \citep{pandas}, 
\texttt{scikit-learn} \citep{sklearn}, 
\texttt{CMasher} \citep{cmasher}}

\bibliographystyle{aasjournal}
\bibliography{Asteroseismology_of_Metal_Poor_Red_Giants_Observed_by_TESS_arxiv.bbl}

\end{document}